\documentclass[pdftex, aps, prd, letterpaper, superscriptaddress, nofootinbib, twocolumn,
               notitlepage, preprintnumbers, longbibliography]{revtex4-1}
\pdfoutput=1
\usepackage{amsmath,amssymb,amsbsy}
\usepackage[utf8]{inputenc}
\usepackage{graphicx,booktabs,longtable,tabularx,capt-of}
\usepackage{soul,color,topcapt}
\usepackage[colorlinks=true, linkcolor=blue, filecolor=blue, urlcolor=blue, citecolor=blue, pdftex, plainpages=false]{hyperref}

\definecolor{orange}{rgb}{1.0, 0.5, 0}

\newcommand{\MSbar}{\ensuremath{\overline{\textrm{MS}} } }

\newcommand{\vev}[1]{\ensuremath{\left\langle #1 \right\rangle} }

\begin{document}
\title{Gradient flow step-scaling function for SU(3) with \texorpdfstring{$N_f=8$}{Nf=8} fundamental flavors}
\author{Anna Hasenfratz}
\email{anna.hasenfratz@colorado.edu}
\affiliation{Department of Physics, University of Colorado, Boulder, Colorado 80309, USA}
\author{Claudio Rebbi}
\affiliation{Department of Physics and Center for Computational Science, Boston University, Boston, Massachusetts 02215, USA}
\author{Oliver Witzel}
\email{oliver.witzel@uni-siegen.de}
\affiliation{Center for Particle Physics Siegen, Theoretische Physik 1, Naturwissenschaftlich-Technische Fakult\"at,
  Universit\"at Siegen, 57068 Siegen, Germany}

\preprint{FERMILAB-PUB-22-785-V,~~SI-HEP-2022-31}

\date{\today}

\begin{abstract}
 The step-scaling function, the lattice analog of the renormalization group $\beta$ function, is presented for the SU(3) gauge system with eight flavors in the fundamental representation. Our investigation is based on generating dynamical eight flavor gauge field configurations using stout-smeared Möbius domain wall fermions and Symanzik gauge action. On these gauge field configurations we perform gradient flow measurements using the Zeuthen, Wilson, or Symanzik kernel and consider the Symanzik, Wilson plaquette, or clover operators to determine step-scaling functions for a scale change $s=2$ including large, up to $48^4$, volumes. Considering different flows and operators as well as the optional use of tree-level improvement allows us to check for possible systematic effects. Our result covers the range of renormalized coupling up to $g_c^2 \lesssim 10$. In the case of $N_f=8$ we observe that the reach in $g_c^2$ is limited due to an unphysical first order bulk phase transition caused by large ultra-violet fluctuations. 

We compare our findings to $N_f=4$, 6, 10 or 12 flavors results that are obtained using the same lattice action and analysis. In addition we investigate the phase structure for simulations with different number of flavors using stout-smeared Möbius domain wall fermions and Symanzik gauge actions to shed some light on the limited reach in $g_c^2$.
\end{abstract}
\maketitle
\section{Introduction}

The SU(3) gauge  theory with  $N_f=8$ fundamental fermions is among the most interesting beyond quantum chromodynamics (QCD) systems. Even though it has been studied in lattice simulations extensively, its infrared nature, i.e.~whether it is conformal or chirally broken, is still unknown (see e.g.~\cite{Appelquist:2007hu,Deuzeman:2008sc,Fodor:2009wk,Hasenfratz:2014rna,Fodor:2015baa,Schaich:2015psa,Appelquist:2014zsa,Aoki:2013xza,Aoki:2014xpa,Aoki:2016wnc,Appelquist:2016viq,Appelquist:2018yqe,Kotov:2021mgp} and references therein).
It has even been suggested that due to special anomaly cancellations in the massless model, $N_f=8$ flavors might be the  sill of the conformal window \cite{Hasenfratz:2022qan}. 

In any case, the $N_f=8$ system is expected to be close to the conformal window, making it an excellent choice  for composite Higgs models, either with all eight flavors massless or as a mass-split system \cite{Hasenfratz:2015xca,Brower:2015owo,Hasenfratz:2016gut,Hasenfratz:2016uar,Appelquist:2020xua,Witzel:2020hyr} where some of the flavors are ``heavy" and decouple in the infrared (IR) limit. In  applications  like the composite Higgs model, it is assumed that the system is chirally broken in the infrared (IR), but a ``nearby"  infrared fixed point (IRFP) drives its low-energy dynamics. Such an IRFP occurs at strong coupling where a nonperturbative approach is necessary to study the IR properties of the system. Several lattice groups have carried out large scale simulations to investigate the phase structure \cite{Deuzeman:2008sc,Schaich:2015psa,Kotov:2021mgp,Hasenfratz:2022qan}, 
the step scaling renormalization group $\beta$ function \cite{Appelquist:2007hu,Hasenfratz:2014rna,Fodor:2015baa}, and the hadron spectrum \cite{Fodor:2009wk,Appelquist:2014zsa,Aoki:2013xza,Aoki:2014xpa,Aoki:2016wnc,Appelquist:2016viq,Appelquist:2018yqe} of the SU(3) 8-flavor model. While lattice calculations  support the expectation that SU(3) with 8 fundamental fermions is close to the conformal window, even the latest large-scale simulations of the hadron spectrum could not unambiguously determine its infrared nature \cite{Nf8posterFleming}. The analysis of the observed meson spectrum is consistent with a dilaton chiral perturbative description as much as with conformal hyperscaling \cite{Appelquist:2017wcg,LatticeStrongDynamics:2018hun,Appelquist:2019lgk,Golterman:2020tdq,Appelquist:2020bqj}. 

Many of the above mentioned works identified a bulk phase transition  of the $N_f=8$ model that prevented the numerical simulations to investigate the strong coupling regime of the system. 
Recently it was proposed to add a set of Pauli-Villars (PV) style heavy bosons with mass at the cutoff level to remove part of the discretization effects introduced by the fermions \cite{Hasenfratz:2021zsl}. First results for the $N_f=8$ system indicate that the first order bulk phase transition can be weakened and even made continuous once the gauge fields are sufficiently smooth \cite{Hasenfratz:2022qan}. 
Finite size scaling analysis using PV improved actions predict a continuous phase transition favoring a Berezinski, Kosterlitz, Thouless (BKT)  type ``walking" scaling, i.e.~a renormalization group $\beta$ function that just  touches zero \cite{Kaplan:2009kr,Vecchi:2010jz,Gorbenko:2018ncu}.   This scaling behavior  suggests that the 8-flavor system could be the sill where the conformal window opens up. This is an unexpected result that  may have important consequences not only for theories describing beyond standard model physics but also for studies of four dimensional conformal systems in general. The conclusion of ``walking" scaling should be checked by independent lattice studies preferably using different actions and/or different lattice methods.

\begin{figure}[tb]
  \includegraphics[width=\columnwidth]{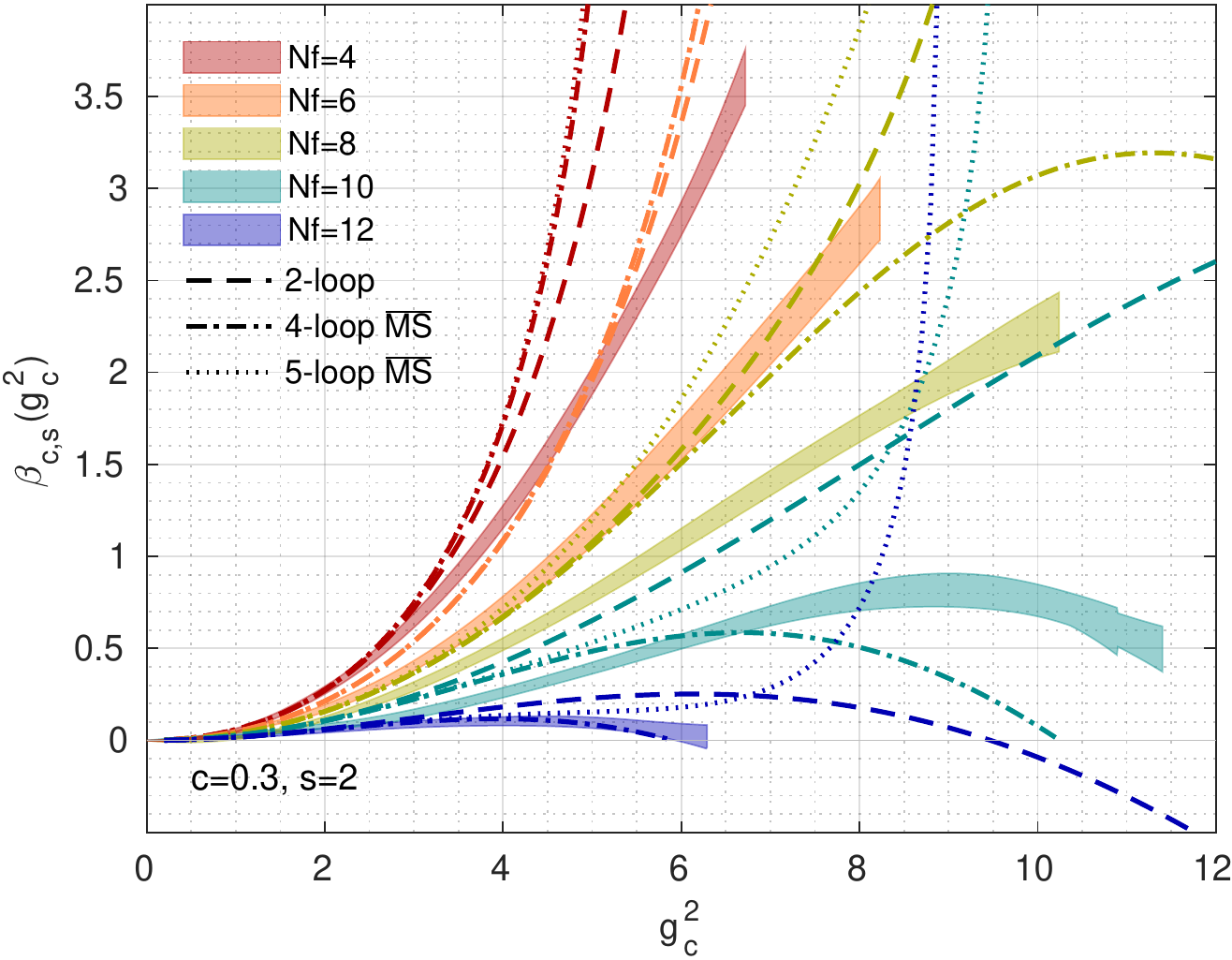}
  \caption{Comparison of the step-scaling functions for $N_f=4$, 6, 8, 10 and 12 using the $c=0.300$ scheme and scale change $s=2$.}
  \label{Fig.All}
\end{figure}

In this work we discuss results on the renormalization group step scaling function of the 8 flavor system using M\"obius domain wall fermions (MDWF). It completes our  systematic investigation of theories with $N_f=2 - 12$ flavors. In previous publications we reported on the $N_f=4$ and 6 \cite{Hasenfratz:2022yws}, 10 \cite{Hasenfratz:2020ess,Hasenfratz:2017qyr}, and 12 \cite{Hasenfratz:2019dpr,Hasenfratz:2017qyr} flavor systems, while we have published results using a slightly different quantity, the continuous $\beta$ function with $N_f=2$ \cite{Hasenfratz:2019hpg,Hasenfratz:2019dpr} and 0 \cite{Peterson:2021lvb} flavors. Figure \ref{Fig.All} summarizes our findings in the $c=0.3$ step-scaling scheme with scale change $s=2$. Comparison of the nonperturbative results with perturbation theory shows that  flavor numbers $N_f\le 8$ run slower than the perturbative predictions. While with 12 flavors there is strong indication of an infrared conformal fixed point, with $N_f=10$ we were not able to reach strong enough couplings to unambiguously identify a fixed point. The reach of our $N_f=8$ simulations is similarly restricted.  The limited range of accessible gauge coupling is due to an unphysical bulk first order phase transitions in lattice simulations. 
Up to renormalized coupling $g^2_{c=0.3}\lesssim 10$ the step scaling function of the 8-flavor system shows a steady rise. This, however, is not in contradiction with the result of Ref.~\cite{Hasenfratz:2022qan} that suggests 8-flavor could be the sill of the conformal window. The predicted value of the gauge coupling at the fixed point is $g^2_{c=0.3} \gtrsim 25$, well outside the reach of the present work.\footnote{Figures 3 and 4 of Ref.~\cite{Hasenfratz:2022qan} show that in the $c=0.45$ scheme $g^2 \approx 30$ at the phase transition. The corresponding value is somewhat smaller with $c=0.30$.} 
Improved lattice actions will be needed to reach stronger gauge couplings in MDWF simulations to be able to verify the claims of staggered fermion simulations in Ref.~\cite{Hasenfratz:2022qan}. 

\begin{figure}[tb]
  \includegraphics[width=\columnwidth]{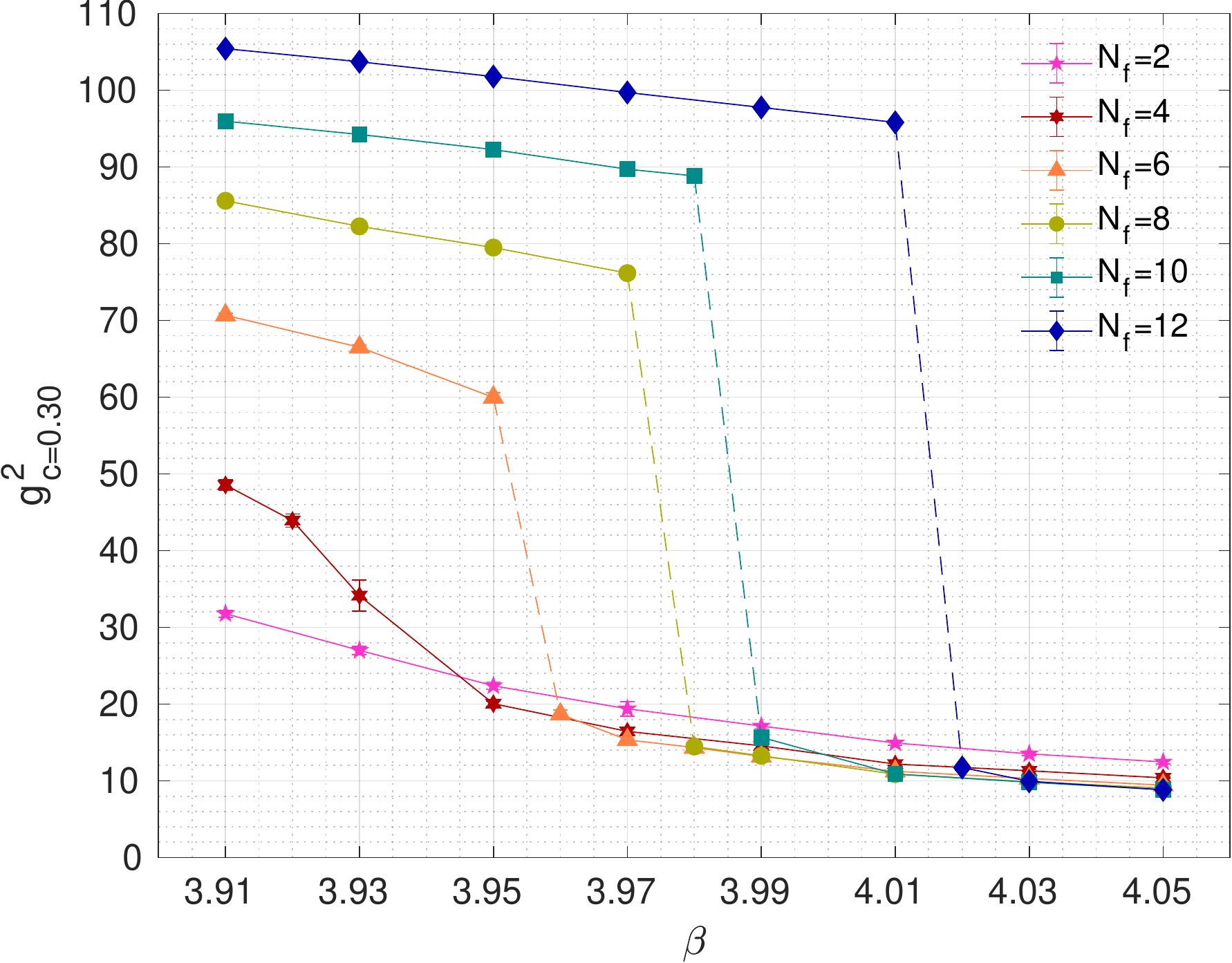}
  \caption{The renormalized gauge coupling determined using Symanzik gauge action, Zeuthen flow, and Symanzik operator in the $c=0.30$ scheme on $8^4$ volumes as the function of the bare coupling for $N_f=2 - 12$ flavors.} 
    \label{Fig.g2}
\end{figure}

The limited reach in the renormalized coupling $g_c^2$ prompted us to study the phase structure of SU(3) gauge system with $N_f=2-12$ flavors in greater detail. Performing dynamical MDWF simulations in the strong coupling region on small $8^4$ lattices, we compute the gradient flow (GF) coupling $g^2_{c=0.3}$ and show how its value varies as we change the bare coupling $\beta$ in Fig.~\ref{Fig.g2}. For systems with six or more flavors, we observe a discontinuity that grows with the flavor number.  This first order transition is at least partially related to lattice ultraviolet fluctuations. For any given lattice action it constrains the bare coupling values that are connected to the perturbative Gaussian fixed point, and consequently limits the largest  renormalized coupling values in finite volumes. With $N_f=8$ flavors the strongest renormalized coupling we can reach is $g^2\approx 14.0$ at $\beta =3.98$. However, the simulations show a wide hysteresis loop that indicates that values with  $\beta < 4.02$ are in a mixed phase. To avoid this problem we 
consider bare couplings $\beta \ge 4.02$. Using lattice volumes $8 \le L/a \le 48$, our  predictions of the $s=2$ step scaling function in the SU(3) 8-flavor system are limited to $g^2_c \lesssim 10$. The accessible range of  renormalized gauge couplings could be increased by using larger volumes, or by improving the lattice action.

In the next section we discuss the details of our lattice setup investigating SU(3) with eight fundamental fermions before we present our step-scaling calculation in Sec.~\ref{Sec.step-scaling}. Subsequently we report further details on our investigations of the  bulk phase transition that restricts the accessible parameter range  with $N_f=2 - 12$ flavors and close by summarizing our work in Sec.~\ref{Sec.Summary}.    
\section{Details of our calculation}
\begin{figure}[tb]
  \includegraphics[width=\columnwidth]{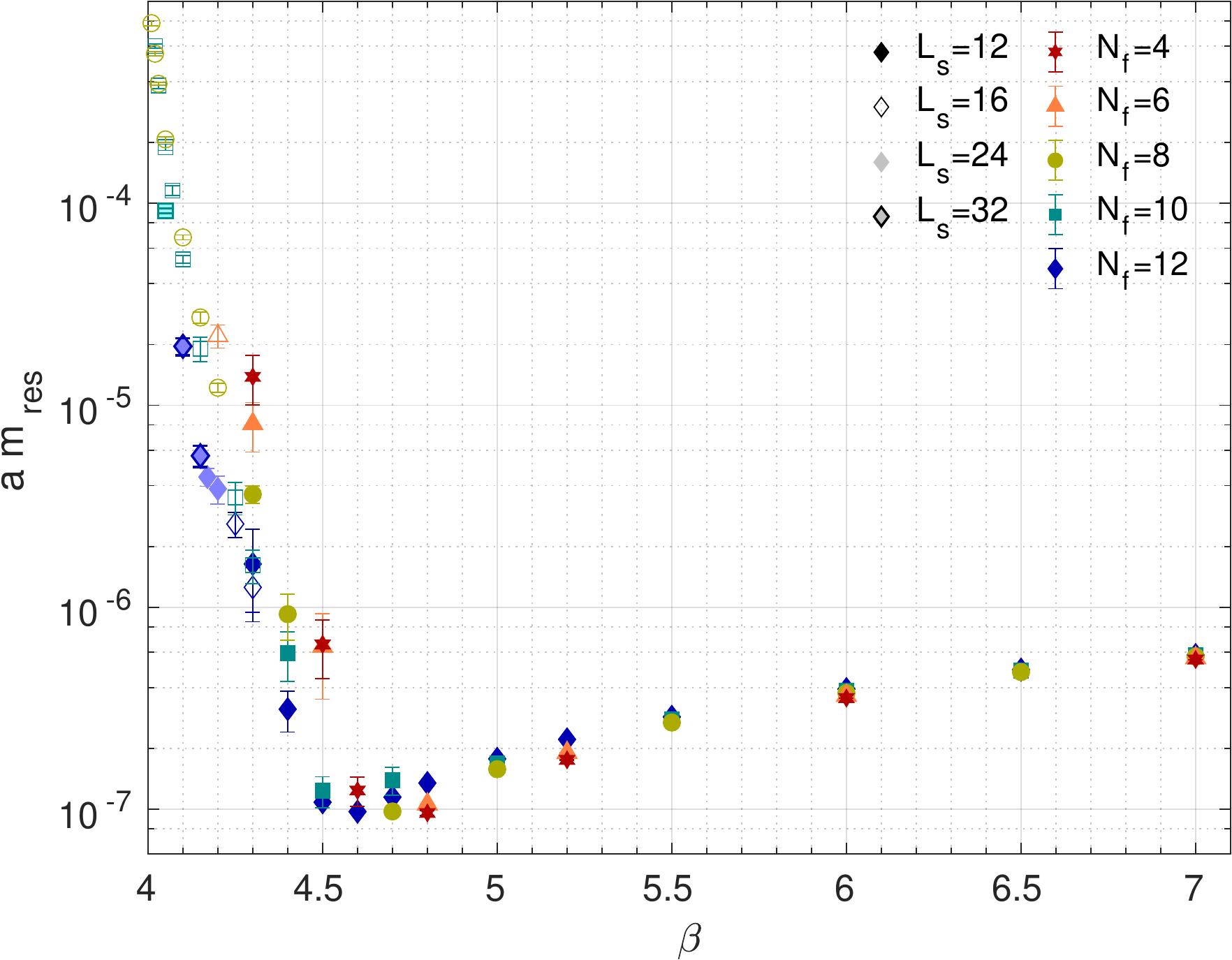}
  \caption{Residual mass $am_\text{res}$ as a function of the bare coupling $\beta$ determined on $(L/a)^4$ volumes with $L/a=32$ for SU(3) gauge systems with $N_f=4$, 6, 8, 10, or 12 flavors. Different colors and symbols distinguish the number of flavors; filled, open, shaded, or framed markers denote the extent $L_s$ of the fifth dimension for MDWF.}
  \label{Fig.Mres}
\end{figure}

We simulate the SU(3) gauge system with eight dynamical fermions in the fundamental representation using the tree-level improved Symanzik (L\"uscher-Weisz) gauge action \cite{Luscher:1984xn,Luscher:1985zq} and three times stout-smeared \cite{Morningstar:2003gk} M\"obius domain wall fermions (MDWF) \cite{Brower:2012vk}. The domain wall height is $M_5=1.0$, the M\"obius parameters are $b_5=1.5$, $c_5=0.5$, and the stout-smearing parameter $\varrho=0.1$. These are the same choices we used for our previous investigations of SU(3) with $N_f=2$ flavors \cite{Hasenfratz:2019hpg}, 4 or 6 flavors \cite{Hasenfratz:2022yws}, 10 \cite{Hasenfratz:2017qyr,Hasenfratz:2020ess} or 12 \cite{Hasenfratz:2017qyr,Hasenfratz:2019dpr} of fundamental fermions. Gauge field configurations are generated with anti-periodic (periodic) boundary conditions for the fermions (gauge field) in all four space-time directions using the hybrid Monte Carlo (HMC) \cite{Duane:1987de} update algorithm as implemented  in \texttt{GRID}\footnote{\url{https://github.com/paboyle/Grid}} \cite{Boyle:2015tjk}. Choosing a trajectory length of $\tau=2$ molecular dynamic time units (MDTU), we save, after thermalization, gauge field configurations every five trajectories. As preferred for step-scaling calculations, we simulate symmetric $(L/a)^4$ hypercubic volumes with antiperiodic boundary conditions for the fermions in all four directions with $L/a=$ 8, 10, 12, 16, 20, 24, 32, and 48 and choose $am_f=0$. Our preferred analysis is based on choosing the scale change $s=2$ considering the five volume pairs $(8\to16)$, $(10\to20)$, $(12\to24)$, $(16\to32)$, and  $(24\to48)$.  For all volumes we perform simulations using bare gauge couplings $\beta\equiv 6/g_0^2 \in\{$7.00, 6.50, 6.00, 5.50, 5.50, 4.70, 4.50, 4.40, 4.30, 4.25, 4.20, 4.20, 4.15, 4.10, 4.05, 4.03, 4.02$\}$, where the smallest $\beta$ values are however only simulated on the smaller volumes to achieve on all $s=2$ volume pairs roughly the same reach in the renormalized coupling and staying in the deconfined regime. The number of generated, thermalized configurations as well as further details are listed in in Table \ref{Tab.Nf8_nZS_ZS} in Appendix \ref{Sec.RenCouplings}. Typically we generated several hundred MDTU on the small volumes, but only 170-200 MDTU on the largest $L/a=48$ volumes. We perform simulations with bare coupling $\beta > 4.20$ using an extent of $L_s=12$ for the fifth dimension of domain wall fermions, while $L_s=16$ is chosen for $\beta\le 4.20$. As demonstrated in our previous work \cite{Hasenfratz:2017qyr,Hasenfratz:2020ess,Hasenfratz:2019dpr} but also shown in Fig.~\ref{Fig.Mres}, this choice ensures that the residual chiral symmetry breaking present for MDWF expressed as the residual mass $am_\text{res}$ remains sufficiently small, below $10^{-4}$ for $\beta \le 4.10$. However, $am_\text{res}$ increases rapidly for even stronger coupling.

Subsequently we read-in these gauge field configurations to perform gradient flow measurements. Gradient flow measurements are separated by 10 MDTU and carried out using \texttt{Qlua}\footnote{\url{https://usqcd.lns.mit.edu/w/index.php/QLUA}} \cite{Pochinsky:2008zz}. We perform a total of three different gradient flows: Wilson (W), Symanzik (S) and Zeuthen (Z) \cite{Ramos:2014kka,Ramos:2015baa} flow. For each flow we determine the Wilson plaquette (W), Symanzik (S) and clover (C) operator to estimate the energy density $\langle E(t)\rangle$ as a function of the gradient flow time $t$. In addition we estimate the topological charge $Q$ at flow time $t$.

While MDWF have in general good chiral properties protecting our zero mass simulations from effects due to nonzero topological charges, we do observe some topological artifacts similar to those encountered in our $N_f=10$ simulations \cite{Hasenfratz:2020vta}. Since statistically only very few artifacts show up within a given set of measurements, we decided to follow the Alpha collaboration \cite{Fritzsch:2013yxa,DallaBrida:2016kgh} and project to the $Q=0$ sector by including only configurations with $|Q|< 0.5$. Using these measurements we perform the statistical data analysis using the $\Gamma$-method \cite{Wolff:2003sm} to estimate and accounts for effects due to autocorrelations.

\section{Step-scaling analysis}
\label{Sec.step-scaling}
Central for the gradient flow step-scaling function, is to define the finite volume gradient flow coupling $g_{GF}^2(t;L,\beta)$ \cite{Fodor:2012td},
\begin{align}
 \label{Eq.g2}
g^2_{GF} (t;L,\beta) = \frac{128\pi^2}{3(N^2 - 1)} \frac{1}{C(t,L/a)} \vev{t^2 E(t)},
\end{align}
where the constants are chosen to match the perturbative 1-loop result in the \MSbar scheme \cite{Luscher:2010iy} with $N=3$ for the SU(3) gauge group.  The coefficient $C(t,L/a)$ is a perturbatively computed tree-level improvement term\footnote{Numerical values for $L/a\le 32$ are listed Table III of Ref.~\cite{Hasenfratz:2019dpr} and for $L/a>32$ in Table V of Ref.~\cite{Hasenfratz:2022yws}.} \cite{Fodor:2014cpa}. When we analyze the data without tree-level improvement, we compensate for zero modes of the gauge field in periodic volumes by replacing $C(c,L/a)$ with $1/(1+\delta(t/L^2))$ \cite{Fodor:2012td}. The flow time $t$ is connected to the lattice size $L$,
\begin{align}
  t=(c L)^2/8,
  \label{Eq.RenCon}
\end{align}
and the parameter $c$ specifies the finite volume renormalization scheme. In order to obtain the gradient flow step-scaling $\beta$ function \cite{Fodor:2012td} for a scale change $s$, the difference of the gradient flow coupling on volume $(L/a)^4$ and $(s\cdot L/a)^4$ needs to be determined
\begin{align}
  \beta_{c,s}(g^2_c;L,\beta) = \frac{g^2_c(sL; \beta)- g^2_c(L; \beta)}{\log\;s^2} \,,
  \label{Eq.beta_cs}
\end{align}
with $g_c^2(L,\beta) =g^2_{GF}(t=(c L)^2/8; L,\beta)$. Defining the re\-normalized coupling $g^2_c$ at a bare coupling $\beta$ implies that $g_c^2$ is subject to cutoff effects. The phenomenologically meaningful result is obtained after taking the continuum limit, which for the step-scaling function corresponds to taking $t/a^2 \to \infty$, or equivalently $L/a \to \infty$ while keeping $g^2_c(L; \beta)$ fixed. Thus at a fixed value of $g^2_c$, the bare coupling is tuned toward the Gaussian fixed point i.e.~$g_0^2= 6/ \beta \to 0$ for increasing $L/a$. In practice we perform simulations on a limited set of lattice volumes and compensate for that by simulating at different values of the bare coupling $\beta$. Combining these simulations at different bare coupling, allows us to cover the investigated range of the renormalized coupling and enables to take the $L/a\to \infty$ continuum limit of the step-scaling $\beta_{c,s}(g^2_c;L)$ at fixed $g_c^2$. In the end this leads to the continuum step-scaling $\beta$-function $\beta_{c,s}(g_c^2)$ in the renormalization scheme $c$.\\

Our analysis starts by following Eq.~(\ref{Eq.g2}) to calculate renormalized couplings $g_c^2(L,\beta)$ for all volumes using a given flow-operator combination (with or without tree-level improvement) and either of the three renormalization schemes ($c=0.300$, 0.275, and 0.250) considered. In the following we refer to the different flow and operator combinations using the shorthand notation [flow][operator] (indicated by the first capital letter) and prefix an ``n'' when the tree-level improvement term $C(c,L/a)$ is included in our analysis. As we detail later, our preferred analysis is based on Zeuthen flow and Symanzik operator, both with and without the use of tree-level improvement and referred to as (n)ZS. For these (n)ZS combinations we list the renormalized couplings together with corresponding integrated autocorrelation times in Table \ref{Tab.Nf8_nZS_ZS} in Appendix \ref{Sec.RenCouplings} and will use (n)ZS in the following to detail our analysis steps.

\begin{figure*}[t]
  \begin{minipage}{0.49\textwidth}
   \flushright 
   \includegraphics[width=0.96\textwidth]{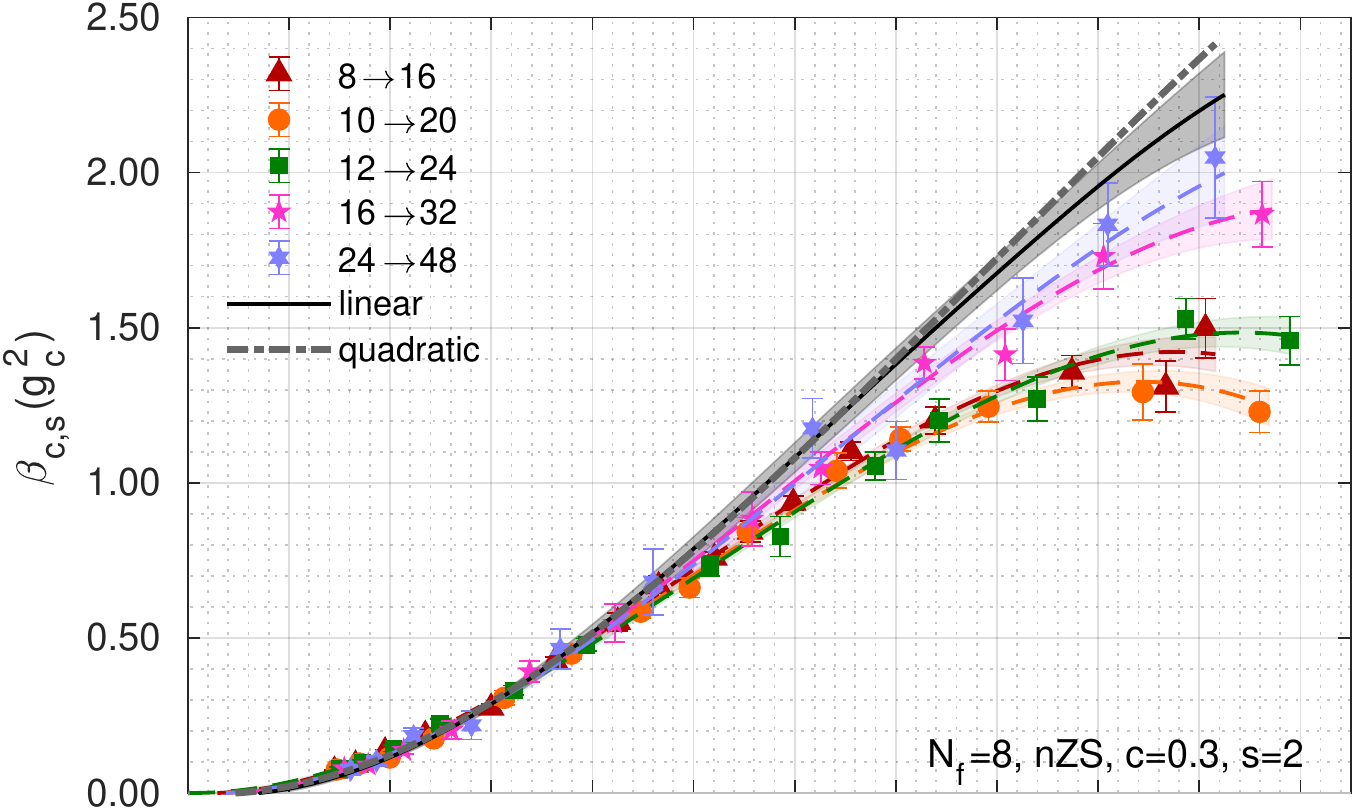}\\
   \includegraphics[width=0.932\textwidth]{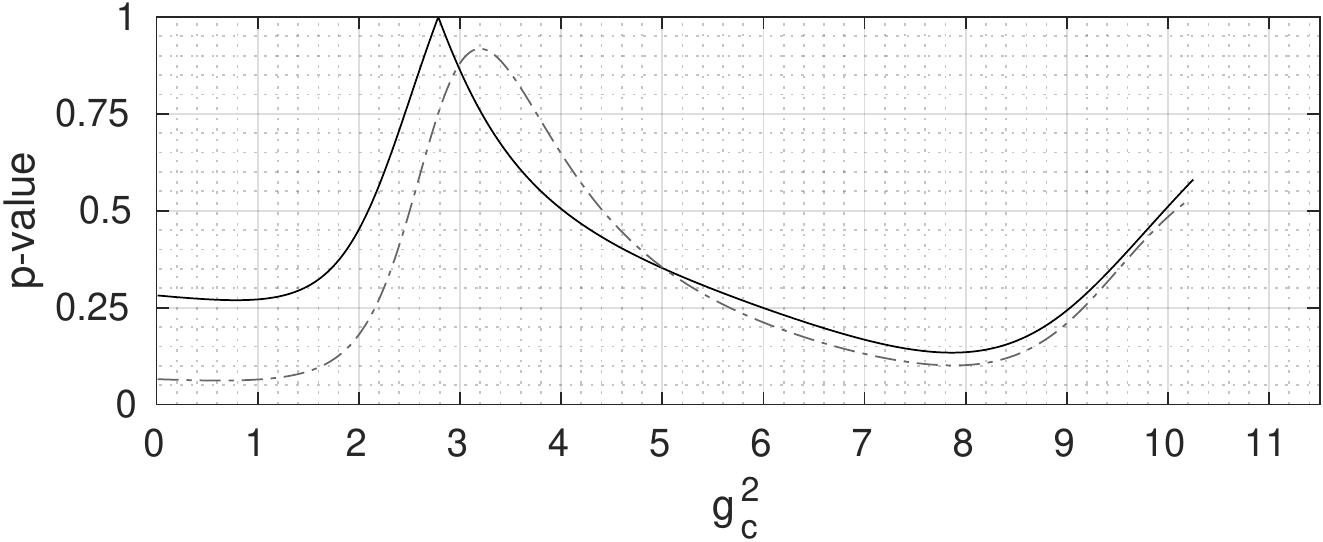} 
   \includegraphics[width=0.96\textwidth]{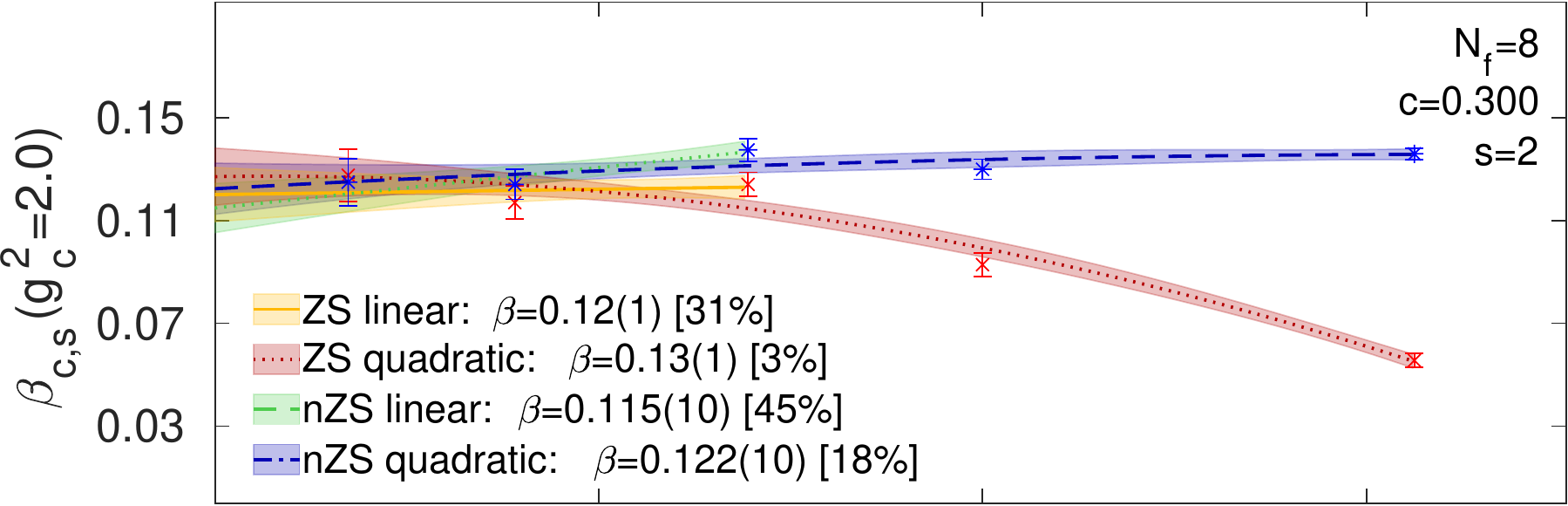}\\
   \includegraphics[width=0.96\textwidth]{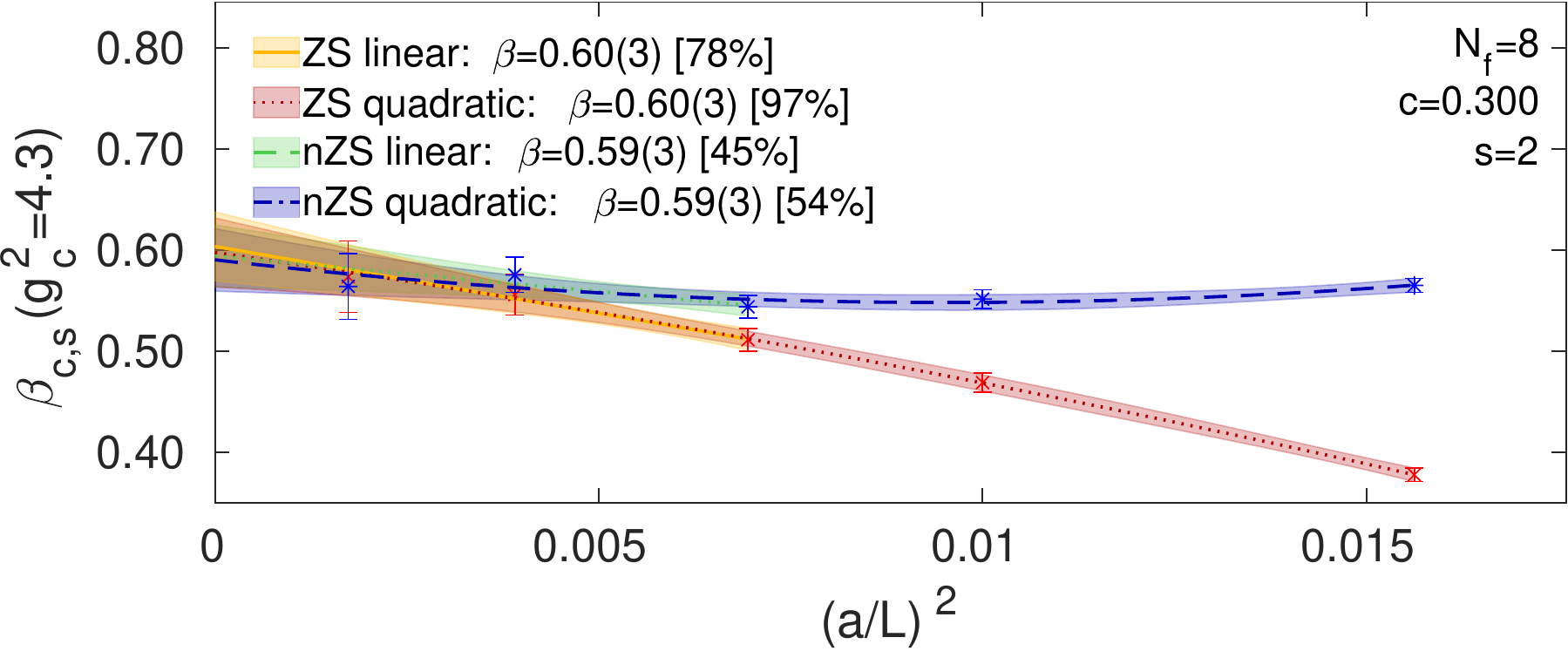}
  \end{minipage}
  \begin{minipage}{0.49\textwidth}
    \flushright
    \includegraphics[width=0.96\textwidth]{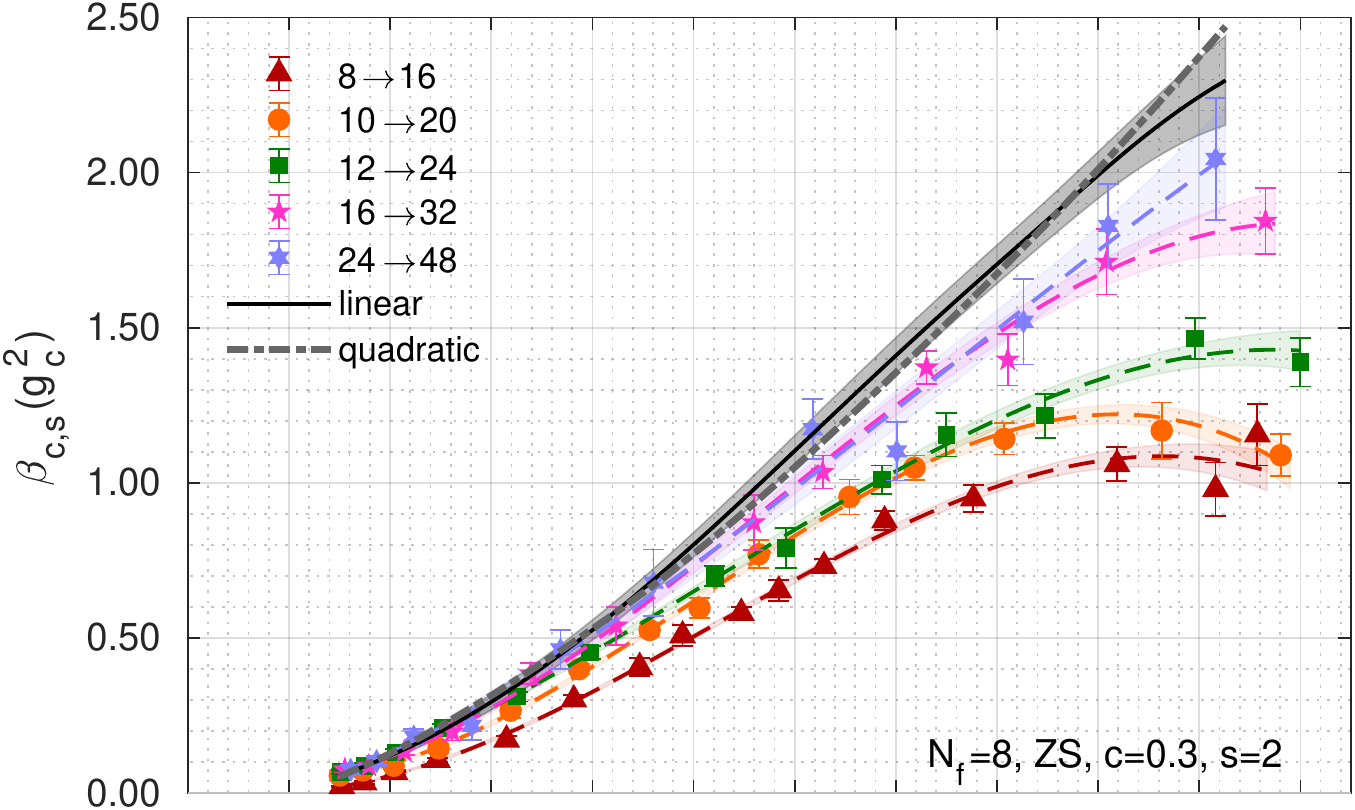}\\    
    \includegraphics[width=0.932\textwidth]{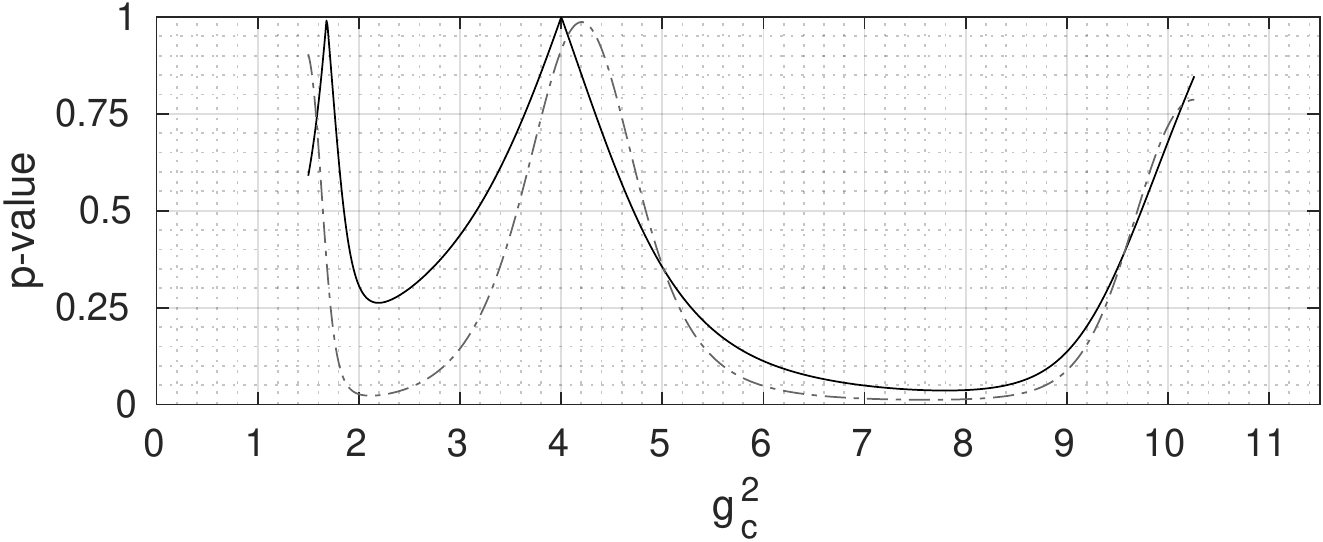} 
    \includegraphics[width=0.96\textwidth]{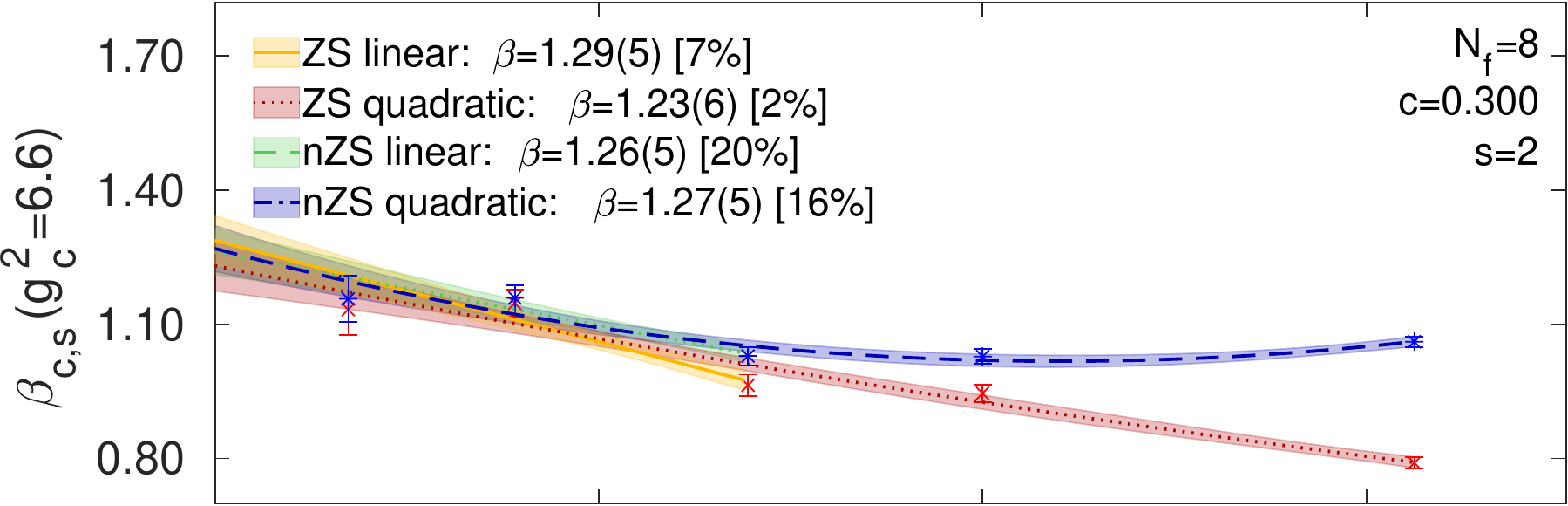}\\
    \includegraphics[width=0.96\textwidth]{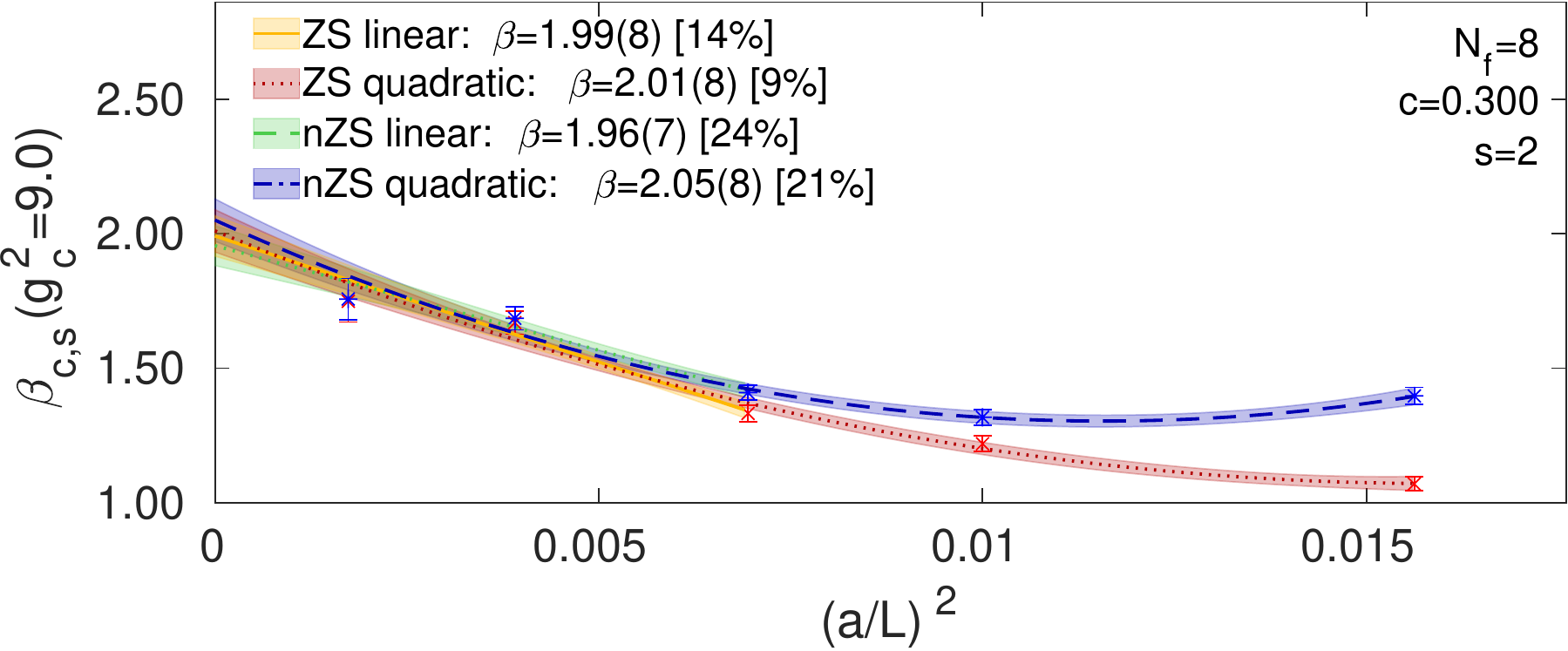}
  \end{minipage}
  \caption{Discrete step-scaling $\beta$-function for $N_f=8$ in the $c=0.300$ gradient flow scheme for our preferred nZS (left) and ZS (right) data sets. The symbols in the top row show our results for the finite volume discrete $\beta$ function with scale change $s=2$. The dashed lines with shaded error bands in the same color of the data points show the interpolating fits. We consider two continuum limits: a linear fit (black line with gray error band) in $a^2/L^2$ to the three largest volume pairs and a quadratic fit to all volume pairs (black dash-dotted line). The $p$-values of the continuum extrapolation fits are shown in the plots in the second row. Further details of the continuum extrapolation at selected $g_c^2$ values are presented in the small panels at the bottom where the legend lists the extrapolated values in the continuum limit with $p$-values in brackets. Only statistical errors are shown.}
  \label{Fig.Nf8_beta_c300}
\end{figure*}

Next we calculate discrete $\beta_{c,s}(g_c^2;L)$ functions, defined in Eq.~(\ref{Eq.beta_cs}), for all five volume pairs with scale change of $s=2$. We show these discrete $\beta_{c,s}(g_c^2;L)$ functions by the colored symbols in the top row plots in Fig.~\ref{Fig.Nf8_beta_c300}.  Figure \ref{Fig.Nf8_beta_c300} shows our analysis for the $c=0.300$ renormalization scheme and corresponding plots for schemes $c=0.275$ and $0.250$ are shown in the Appendix \ref{Sec.c0275_c0250}, Figs.~\ref{Fig.Nf8_beta_c275} and \ref{Fig.Nf8_beta_c250}, respectively.

Motivated by the perturbative expansion
  \begin{align}
    \beta_{c,s}(g_c^2;L) = \sum_{i=0}^{n} b_i g_c^{2i}.
    \label{Eq.fit_form}
  \end{align}
  we interpolate these discrete $\beta_{c,s}(g_c^2;L)$ functions by performing a polynomial fit and achieve a good description of our data using a polynomial of degree $n=3$. Since discretization effects at weak coupling are sufficiently small when using tree-level normalization (tln), we constrain the intercept $b_0$ to vanish but fit $b_0$ without tln. The outcome of these interpolating fits are listed in Tab.~\ref{Tab.interpolationsNf8} and the resulting finite volume discrete step-scaling functions $\beta_{c,s}(g_c^2; L)$ at continuous values of $g_c^2$ are shown in top row plots of Figs.~\ref{Fig.Nf8_beta_c300}, \ref{Fig.Nf8_beta_c275}, and \ref{Fig.Nf8_beta_c250} by the shaded bands in the same color as the values of the discrete $\beta_{c,s}(g_c^2;L)$ functions.

In the next step we extrapolate these continuous-in-$g_c^2$ finite volume discrete step-scaling functions to the infinite volume continuum limit at fixed values of $g^2_c$ to obtain phenomenologically meaningful results. Specifically we choose two different fit ans\"atze to perform these extrapolations which enables us to check for consistency. Our first choice is to perform a linear fit in $(a/L)^2$ using only our three largest volume pairs $12\to 24$, $16\to 32$, and $24\to 48$. This fit is shown by a solid black line with gray error band in top row plots of Figs.~\ref{Fig.Nf8_beta_c300}, \ref{Fig.Nf8_beta_c275}, and \ref{Fig.Nf8_beta_c250} with corresponding $p$-values given by the solid black in the second row plots. Secondly we perform a quadratic fit in $(a/L)^2$ using all five volume pair and visualize it by the black dash-dotted lines. Details of these continuum extrapolations fits are presented for four selected values of $g_c^2$ in the bottom two rows of Figs.~\ref{Fig.Nf8_beta_c300}, \ref{Fig.Nf8_beta_c275}, and \ref{Fig.Nf8_beta_c250}. While linear and quadratic fits result in consistent continuum step-scaling functions for nZS and ZS for all $c$ schemes across the range of $g_c^2$ covered, the goodness of fit ($p$-value) is typically higher for nZS than ZS. In particular quadratic fits for ZS in the range $5.5\le g_c^2 \le 8.5$ exhibit low, if not zero, $p$-values. Taking a look at the finite volume step-scaling functions in the top row plots, these poor $p$-values correspond to the $8\to 16$ and/or $10\to 20$ data having a different ``shape'' than the other volume pairs. This is a sign of these volumes being too small for these strong coupling. Consequently, we use the linear fits as our preferred analysis and only show quadratic fits for consistency.

\begin{figure*}[p]
  \includegraphics[height=0.30\textheight]{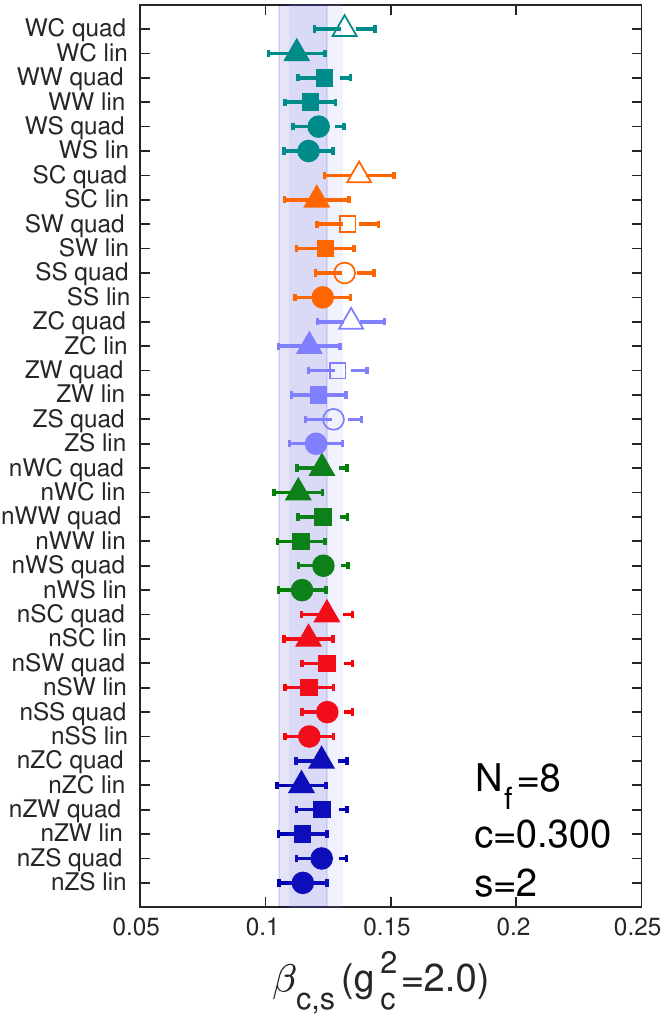}
  \includegraphics[height=0.30\textheight]{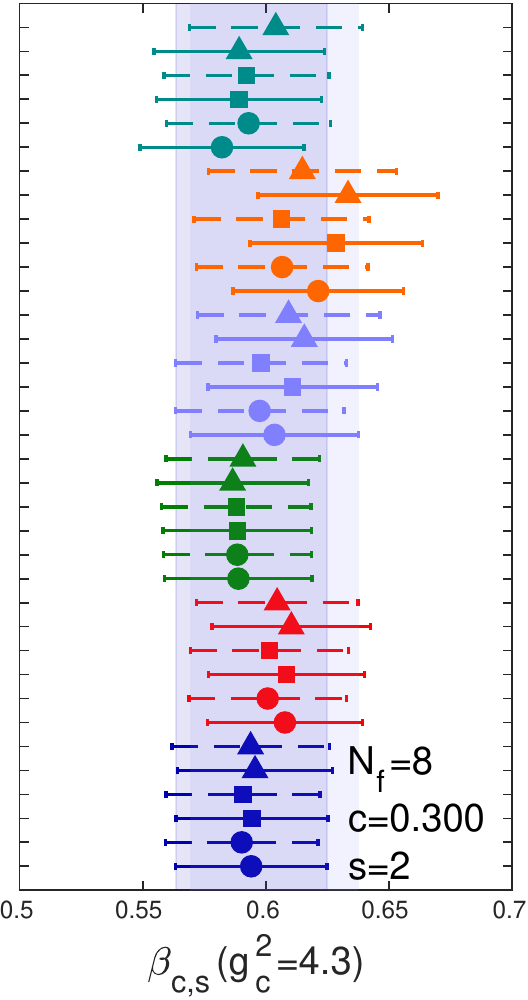}
  \includegraphics[height=0.30\textheight]{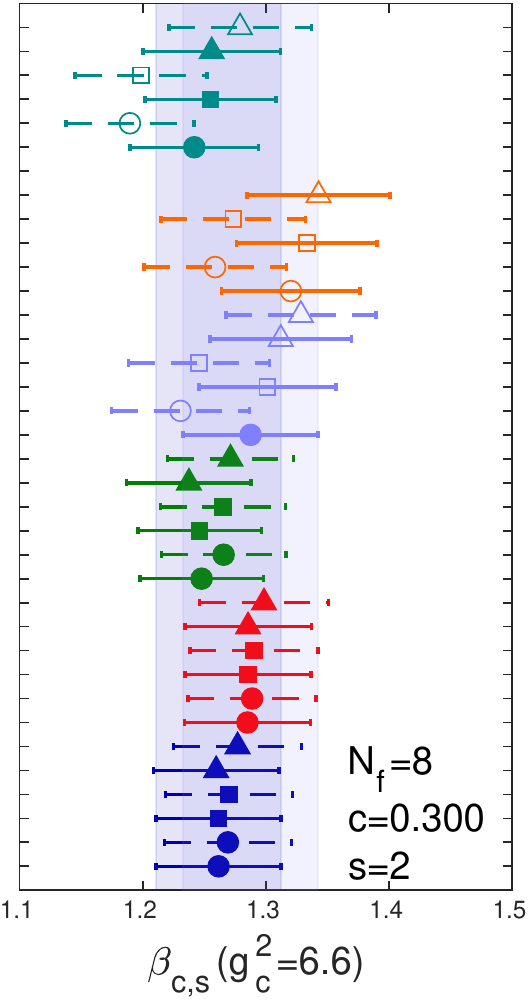}
  \includegraphics[height=0.30\textheight]{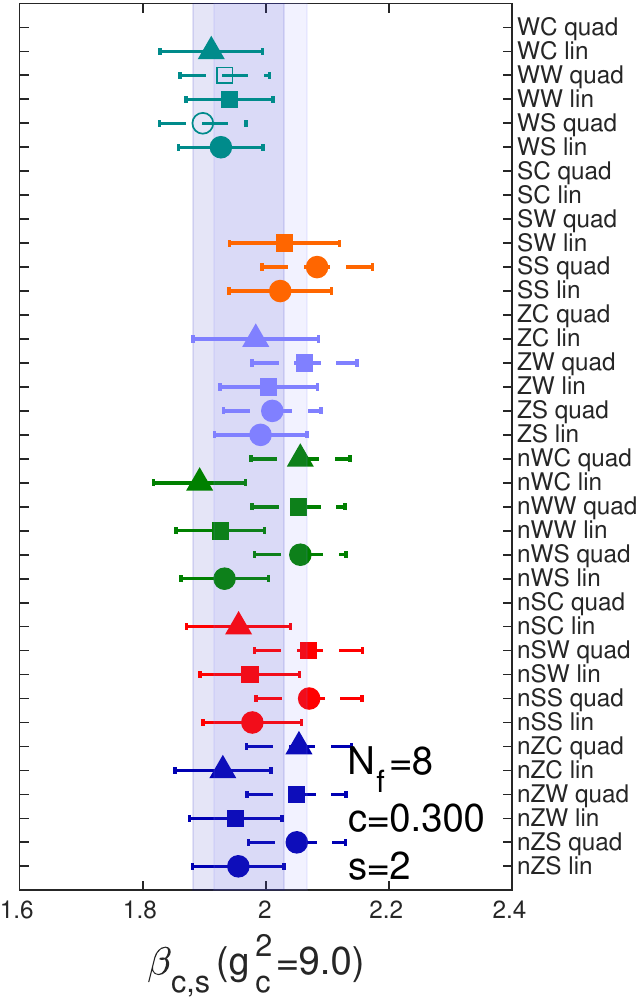}\\  
  \includegraphics[height=0.30\textheight]{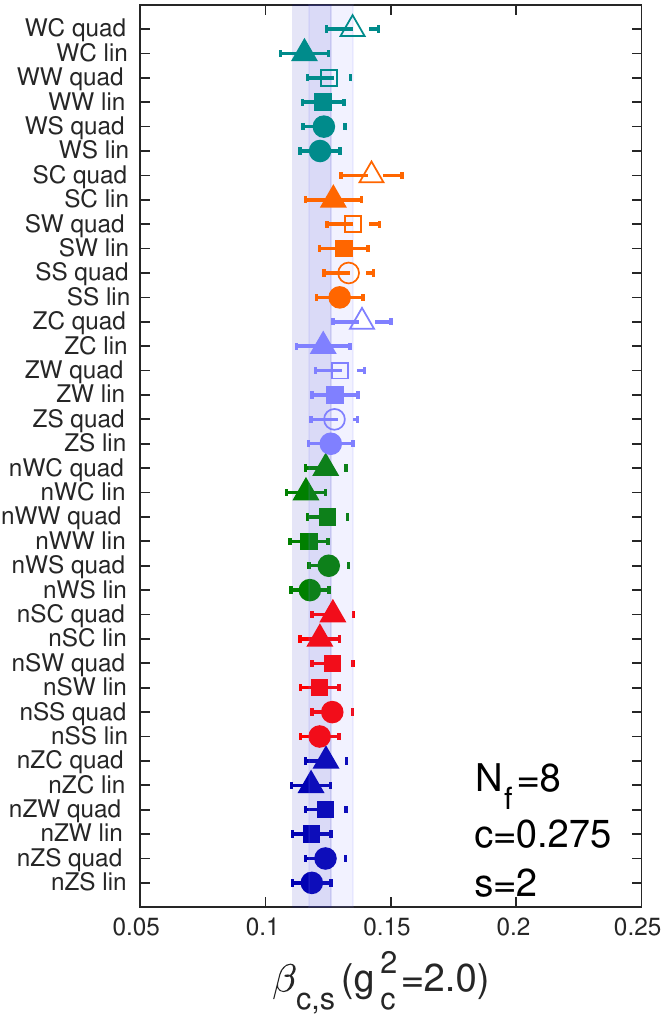}
  \includegraphics[height=0.30\textheight]{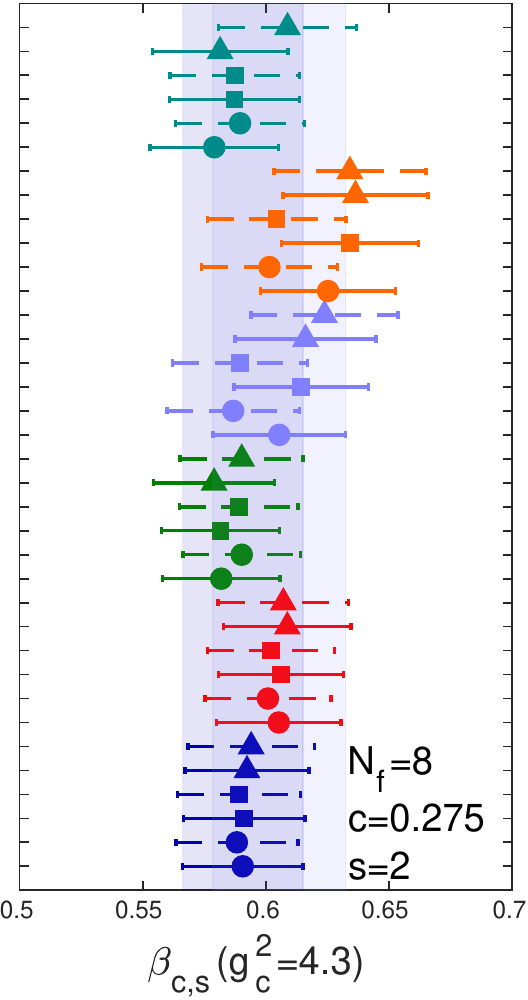}
  \includegraphics[height=0.30\textheight]{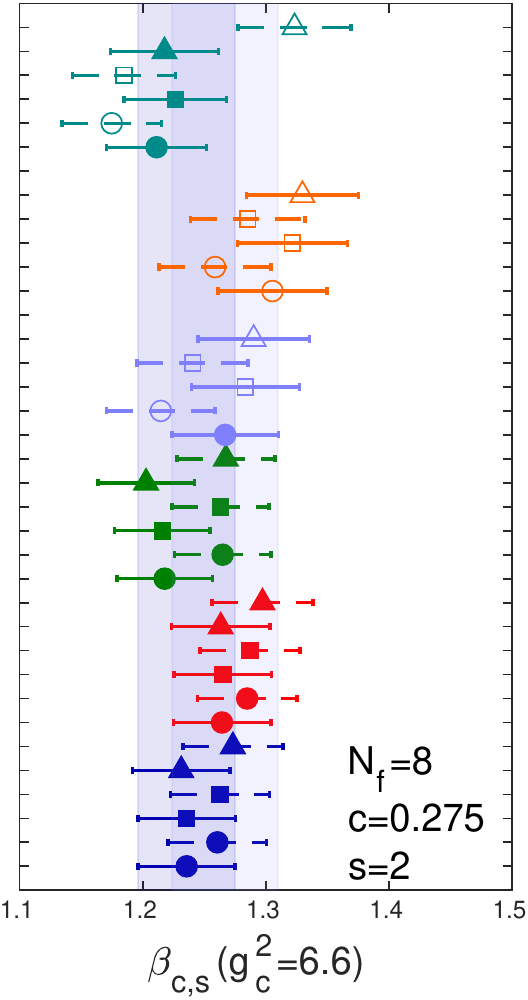}
  \includegraphics[height=0.30\textheight]{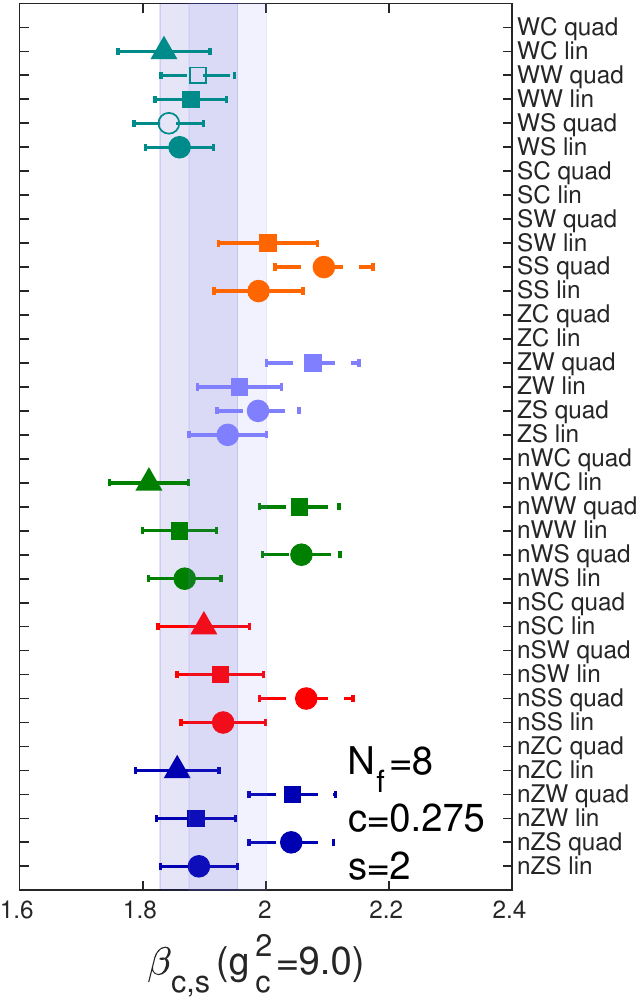}\\
  \includegraphics[height=0.30\textheight]{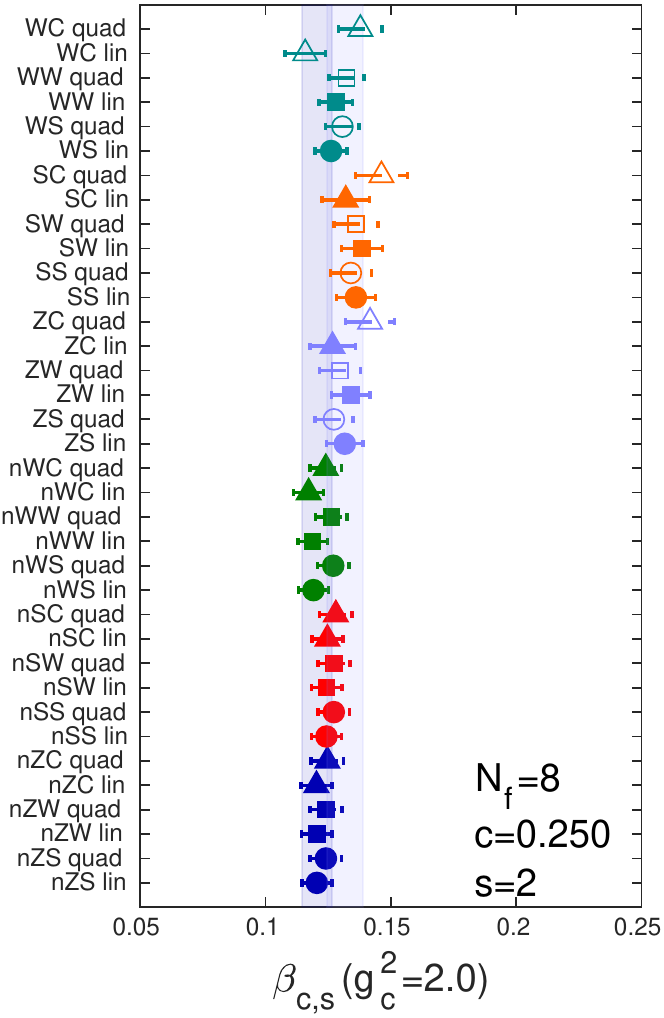}
  \includegraphics[height=0.30\textheight]{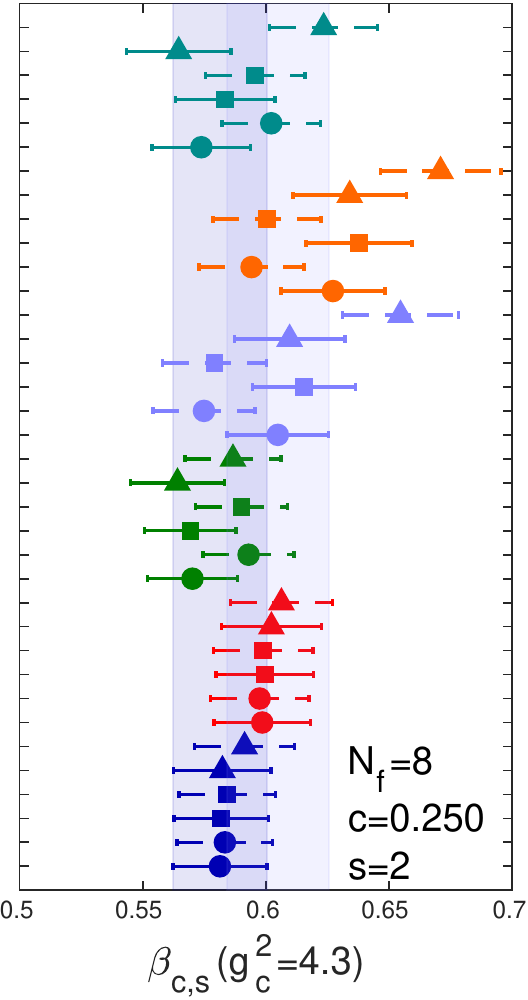}
  \includegraphics[height=0.30\textheight]{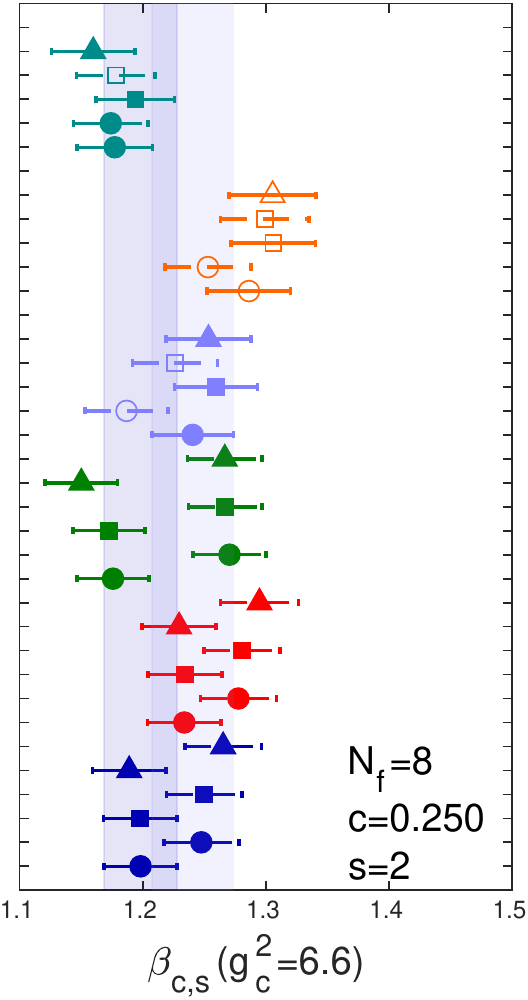}
  \includegraphics[height=0.30\textheight]{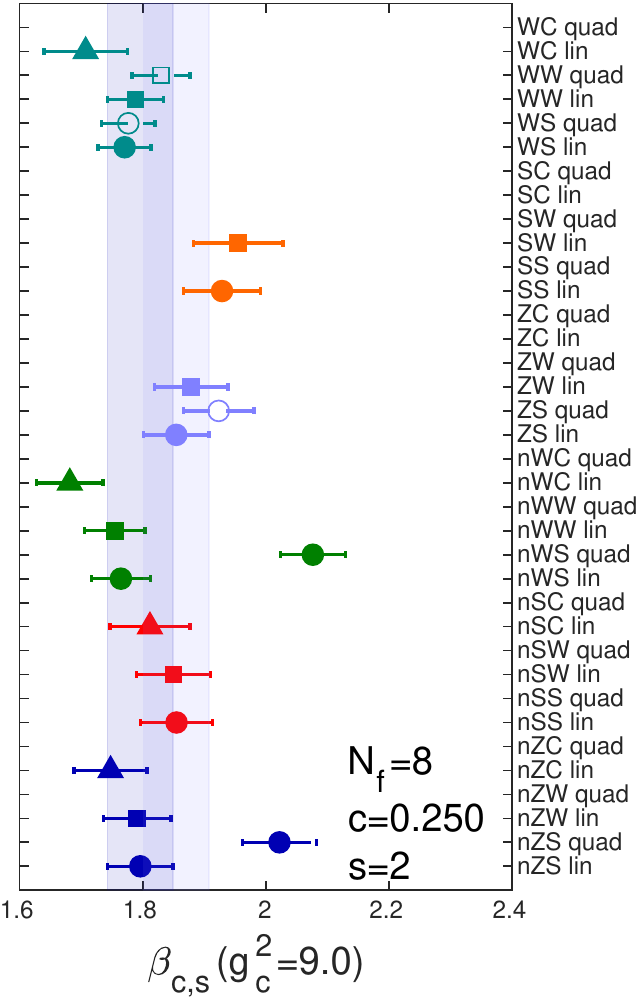}  
  \caption{Systematic effects on the $N_f=8$ results for $\beta_{c,s}(g_c^2)$ due to tree-level improvement, different flows and operators as well as linear or quadratic continuum extrapolation fits. In all cases we obtain the continuum limit considering a linear extrapolation to the three largest volume pairs and a quadratic extrapolation to all volume pairs. The columns show our continuum limit results at selective $g_c^2 = 2.0$, 4.3, 6.6, and 9.0; the rows correspond to renormalization schemes $c=0.300$, 0.275, 0.250. Open symbols indicate extrapolations with a $p$-value below 5\%. The vertical shaded bands highlight our preferred (n)ZS analysis.}
  \label{Fig.Nf8_beta_sys}
\end{figure*}

While the continuum results are expected to be free of discretization effects, they may nevertheless be subject to other systematic effects. In addition to varying the ansatz for the continuum limit extrapolation, we therefore also take advantage of our additional gradient flow measurements and repeat the analysis for all different flow-operator combinations with and without using tln. Choosing again four selected values $g_c^2$ across the range where we have data, we compare the different determinations of $\beta_{c,s}(g_c^2)$ in Fig.~\ref{Fig.Nf8_beta_sys} where the different rows show our three different $c$ schemes and the columns align different $g_c^2$ values. Highlighting our preferred (n)ZS analysis by the shaded blue bands, we observe an overall consistency of the 18 different analysis mostly at the 1$\sigma$ level. However, we note that the spread increases as we move to smaller schemes $c$ and/or to stronger coupling $g_c^2$. The total number of ``outliers'' not touching the blue bands is very small. Therefore we take the envelope of nZS and ZS to obtain our final results which in particular for smaller $c$ values  visibly increases the error of our final result.

We conclude our presentation on the $N_f=8$ step-scaling function by showing how our final, nonperturbative results\footnote{ASCII files containing the data corresponding to our final results (envelope of nZS and ZS) are uploaded as Supplemental Material.} based on nZS + ZS compare to the universal perturbative 1- and 2-loop predictions, the perturbative 3-loop prediction in the gradient flow scheme \cite{Harlander:2016vzb}, and the 3-, 4-, and 5-loop predictions in the $\overline{\textrm{MS}}$ scheme \cite{Baikov:2016tgj,Ryttov:2016ner}.

\begin{figure}[p]
    \includegraphics[width=0.98\columnwidth]{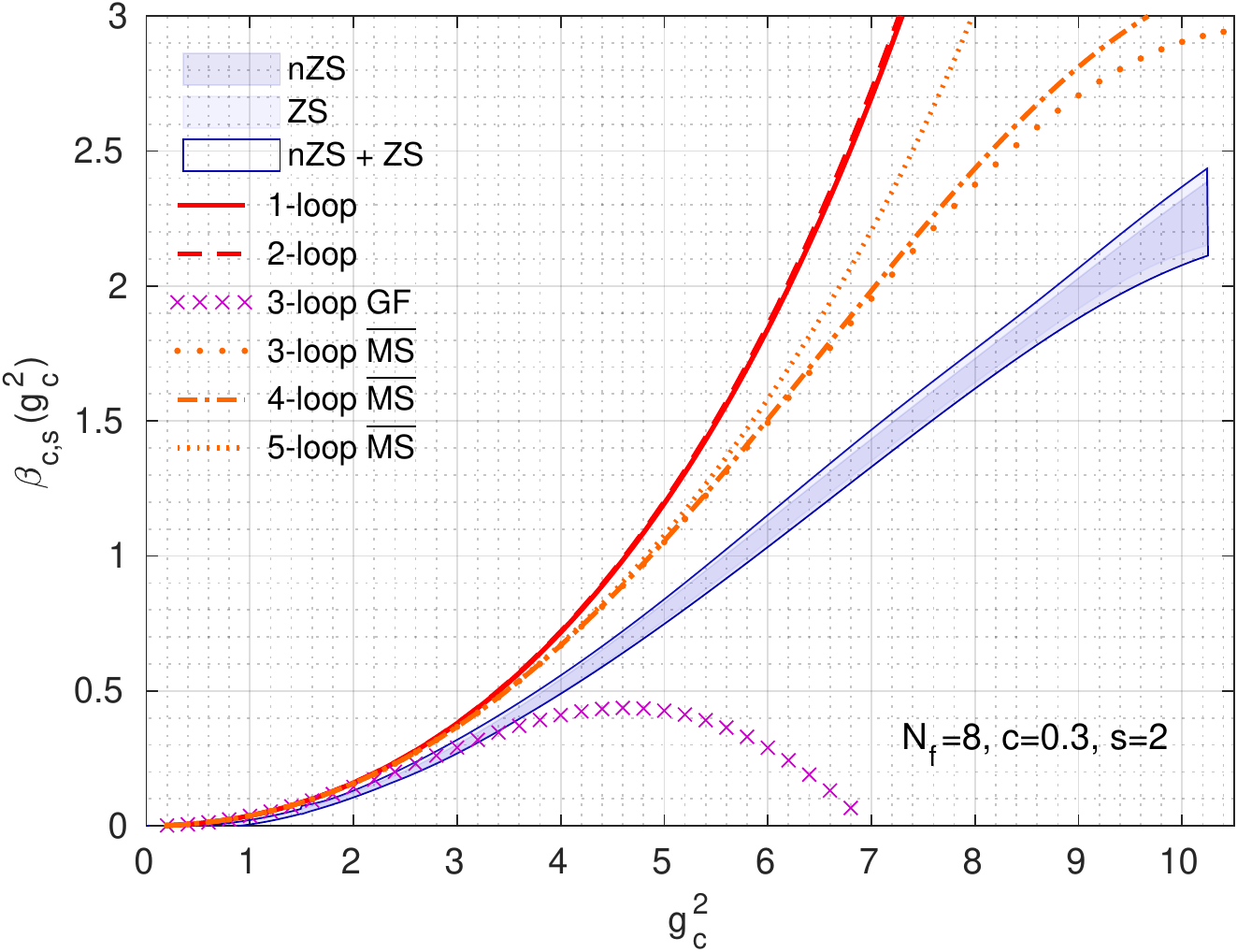}
    \includegraphics[width=0.98\columnwidth]{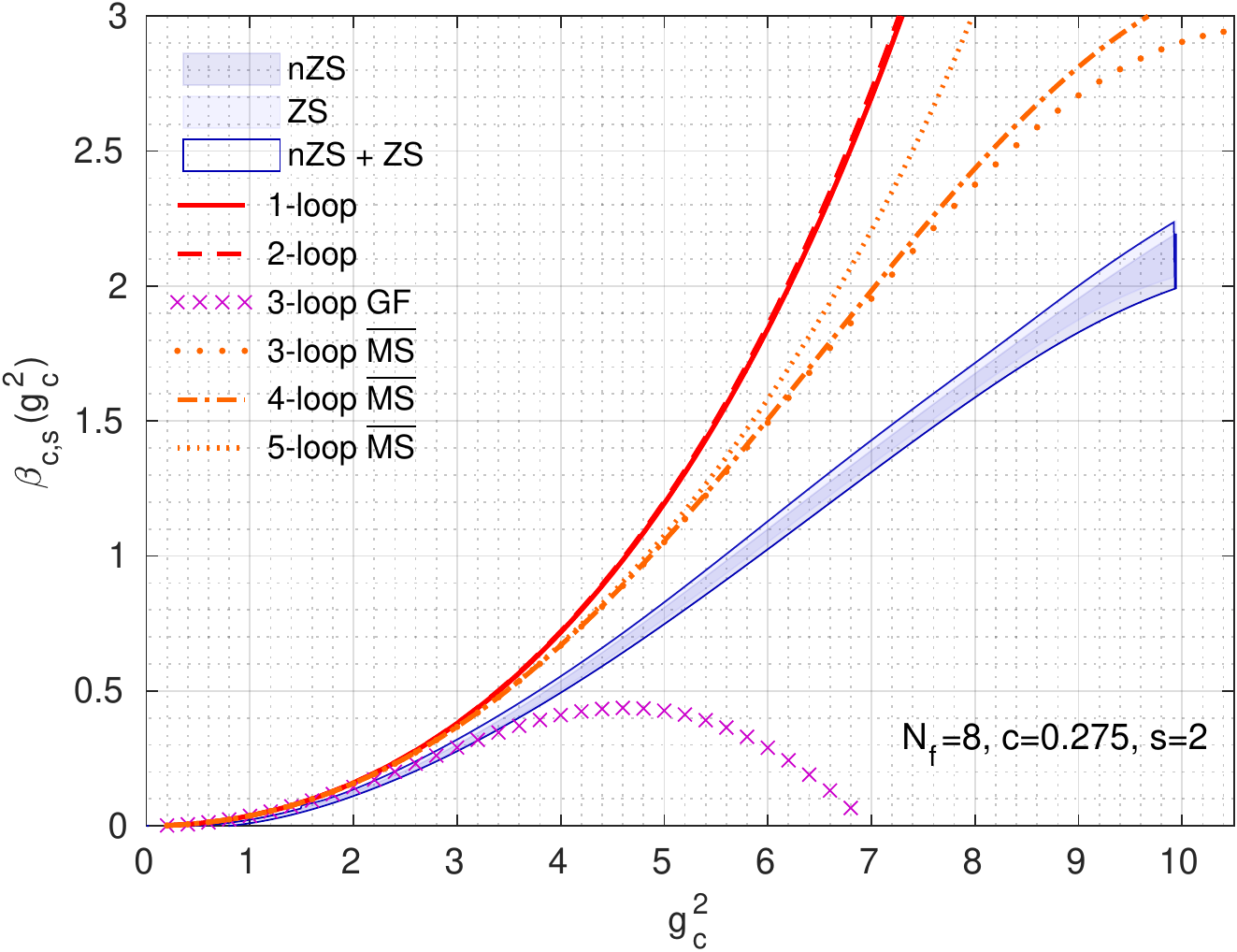} 
    \includegraphics[width=0.98\columnwidth]{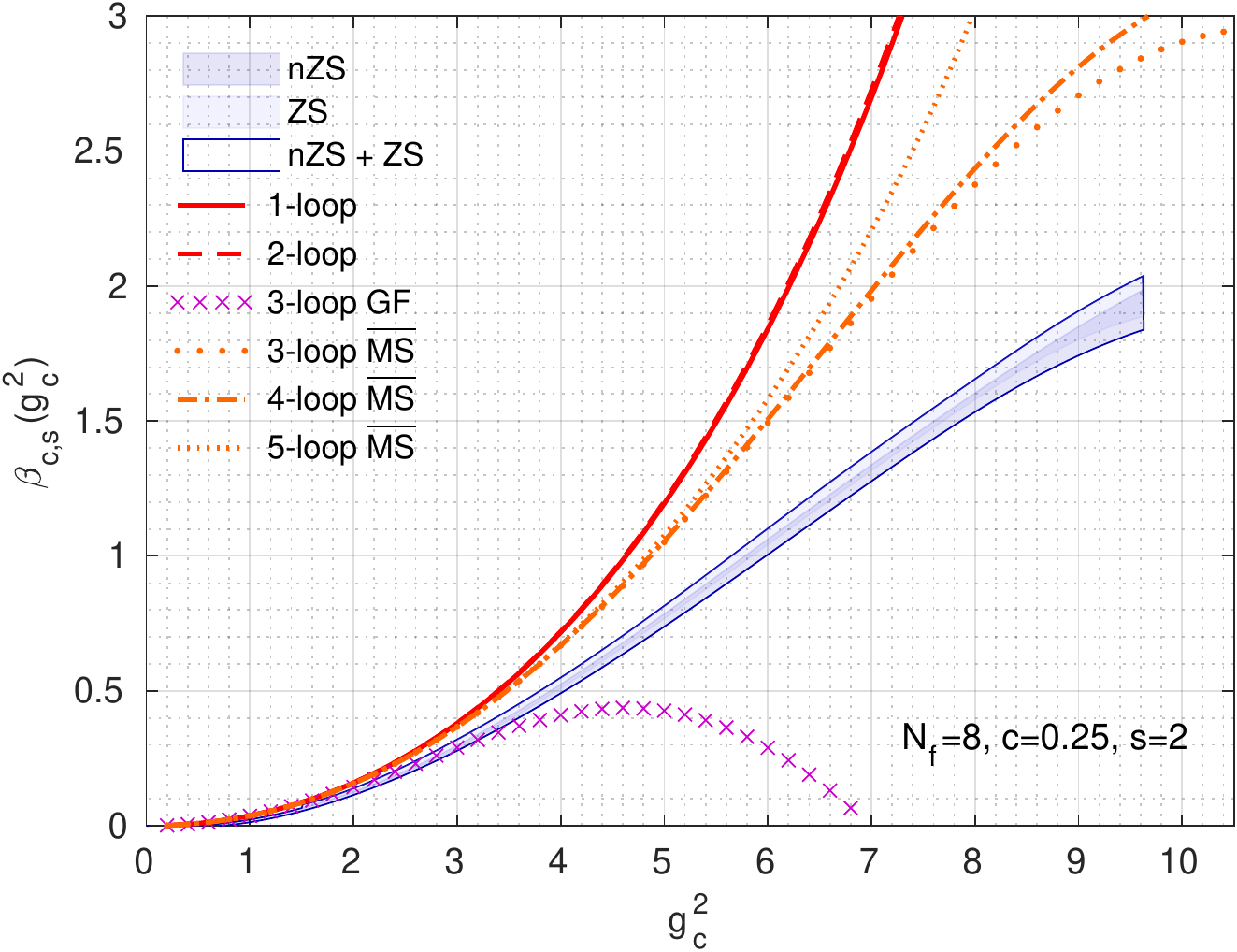}
  \caption{Comparison of our final $N_f=8$ continuum results obtained from our preferred (n)ZS data set for $c = 0.300$ (top), 0.275 (middle), and 0.250 (bottom) to universal 1- and 2-loop perturbative predictions (red), 3-loop perturbative predictions in the gradient flow scheme (purple) and  3-, 4-, and 5-loop $\MSbar$ scheme predictions (orange).}
  \label{Fig.FinalNf8}  
\end{figure}

As in the case of our previous work for $N_f=4$ and 6 flavors \cite{Hasenfratz:2022yws}, we observe that the perturbative step-scaling function runs noticeable slower than the universal or $\overline{\textrm{MS}}$-scheme perturbative predictions. While 1- and 2-loop as well as 3- and 4-loop are very close to each other, the 5-loop prediction does not follow the trend sitting essentially between the two groups. Overall the 3- and 4-loop $\overline{\textrm{MS}}$-scheme predictions are qualitatively closest to our nonperturbative result. The 3-loop gradient flow scheme prediction \cite{Harlander:2016vzb} seems again not to be trustworthy at strong coupling ($g_c^2\gtrsim 4$) because it sharply turns around hinting at a fixed point at $g_c^2\sim 7$ where our nonperturbative $\beta$ function grows steadily. However, for weaker coupling $0\le g_c^2\le 3$ the perturbative 3-loop GF prediction perfectly traces our nonperturbative result. Hence it would be extremely interesting to learn how the perturbative scheme converges when higher loop corrections are considered.

\section{Phase diagram from \texorpdfstring{$N_f= 2 - 12$}{Nf=2-12} with domain wall fermions}
    
Figure \ref{Fig.All} summarizes our results for the $s=2$ step scaling function with $N_f=4$, 6, 8, 10, and 12 flavors. The reach in the renormalized coupling $g^2_c$ is limited by the onset of chiral symmetry breaking for $N_f=2$ and 4 flavors and on larger volumes also for $N_f=6$ flavors. With $N_f\ge 8$ the simulations even on our largest volumes never reach the regime where chiral symmetry is broken, the accessible gauge coupling is limited by the onset of a strong first order bulk phase transition. The situation is similar to staggered fermion simulations where systems with  $N_f\ge 8$ undergo a bulk transition, thus limiting the value of the strongest finite volume GF coupling. In most cases this bulk transition is triggered by strong UV fluctuations and can be mitigated by improving the action. Reference \cite{Hasenfratz:2021zsl} has shown that the inclusion of heavy Pauli-Villars type bosons counter the induced gauge action of the fermions and lead to numerical simulations with smoother gauge fields at identical renormalized gauge coupling. The bulk phase transition caused by UV fluctuations are shifted by the smoother gauge fields and stronger gauge couplings are accessible in simulations. 
Studies of the $N_f=8$ system with staggered fermions and sufficient number of heavy PV bosons suggest that the bulk first order phase transition turns to a bulk continuous phase transition that favors ``walking scaling'', i.e.~a $\beta$ function that just touches zero. This scenario would make $N_f=8$ the sill of the conformal window, a possibility that most likely is related to 't Hooft anomaly cancellation with two sets of staggered fermions \cite{Catterall:2020fep,Butt:2021brl}.
The phase transition occurs at a rather strong $g^2_*$ gauge coupling. The value of $g^2_*$ depends on the renormalization scheme, preliminary results indicate $g^2_* \gtrsim 25 $ in the $c=0.30$  GF scheme and not in the range of existing simulations that do not utilize PV improvement.

Our MDWF simulations has similar limitations as staggered ones. With our action we cannot reach the regime $g^2_c \gtrsim 10$. Trying to push the simulations to stronger coupling we first observe that residual mass $am_\text{res}$, parametrizing the residual chiral symmetry breaking present in domain wall fermions, starts to grow. As we show in Fig.~\ref{Fig.Mres}, the residual mass at weak coupling does not show any dependence on the number of flavors. This changes when the bare coupling drops below 5.5 where slight differences in $am_\text{res}$ for different $N_f$ become visible. These differences grow for stronger coupling likely related to the phase structure of the system. 

To get a better understanding of the 
 phase structure and bulk transitions of SU(3) gauge systems with $N_f=2$, 4, 6, 8, 10, or 12 flavors we performed a large number of dedicated small $8^4$ simulations using the same stout-smeared MDWF with Symanzik gauge action. For all these simulations we fix the fifth extent of domain-wall fermions to be $L_s=12$. 
 First we explore the weak coupling ``branch'' by starting from existing configurations at $\beta=4.05$ and decrease $\beta$ in steps of 0.02 down to 3.91. We observe clear first order phase transitions for $N_f\ge 6$, while $N_f=2$ and 4 show a smooth behavior. Near the transitions  we fill  in steps of 0.01. Second we explore the strong coupling branch starting from configurations at $\beta=3.91$ and increase $\beta$ in steps of 0.02 again filling  in steps of 0.01 near the transitions. For all simulations we generate at least 1000 trajectories with trajectory length $\tau=2$ MDTU and use at least 200 trajectories for thermalization. In cases where the transition occurs ``late'' or we observe interesting fluctuations, we run these ensembles for at least another 1000 trajectories. With our statistics we have not observed multiple tunneling in any of the systems, and in some cases we cannot exclude that a tunneling event may occur later.

\begin{figure}[tb]
  \includegraphics[width=\columnwidth]{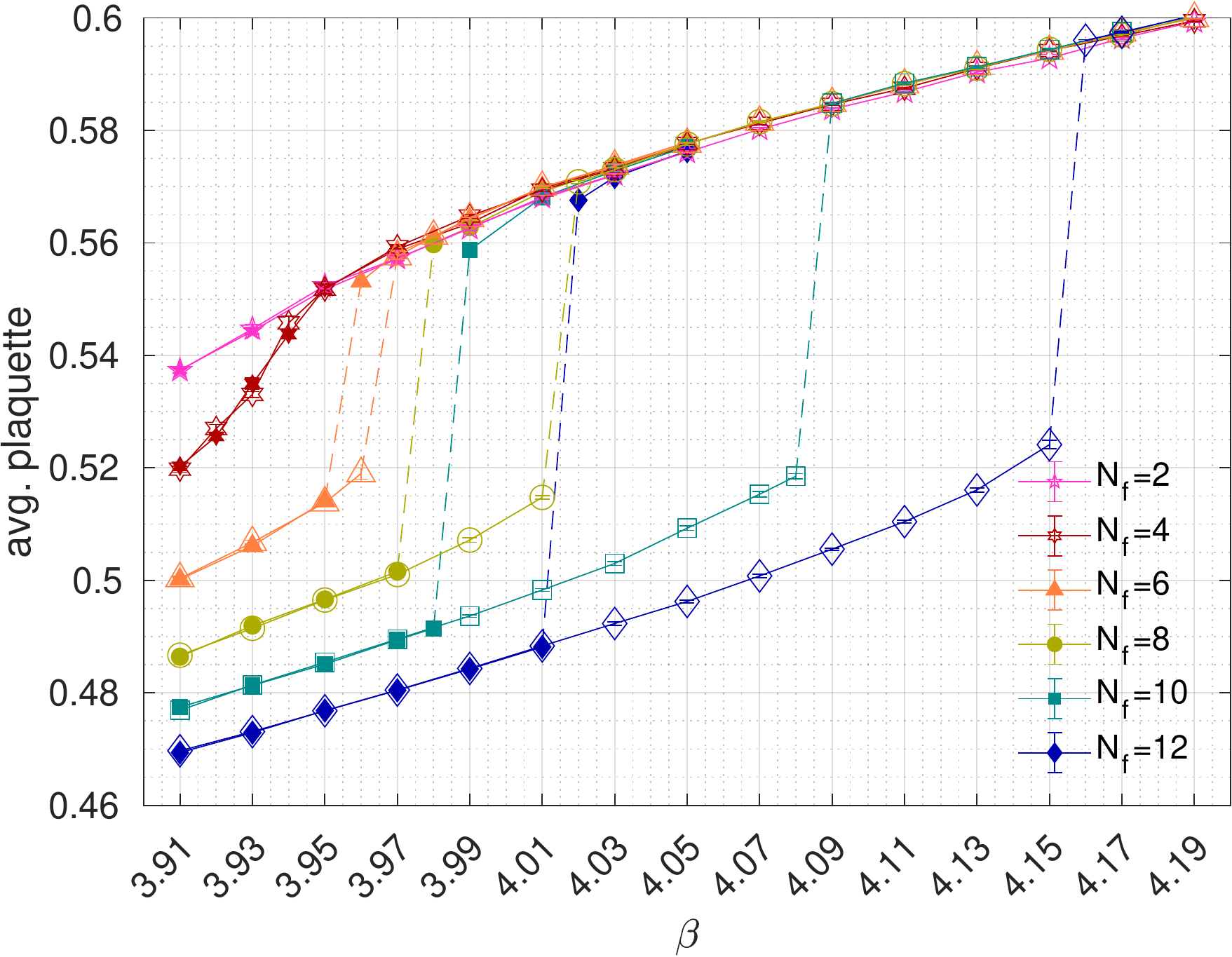}
  \caption{Comparison of the plaquette on $8^4$ volumes as the function of the bare coupling for $N_f = 2 - 12$ flavors. By performing starts from weak and strong coupling we are resolving a hysteresis for $N_f\ge 6$. Both the discontinuity and the hysteresis width grows with increasing flavor number.}
  \label{Fig.Plaquette}
\end{figure}

\begin{figure}[tb]
  \includegraphics[width=\columnwidth]{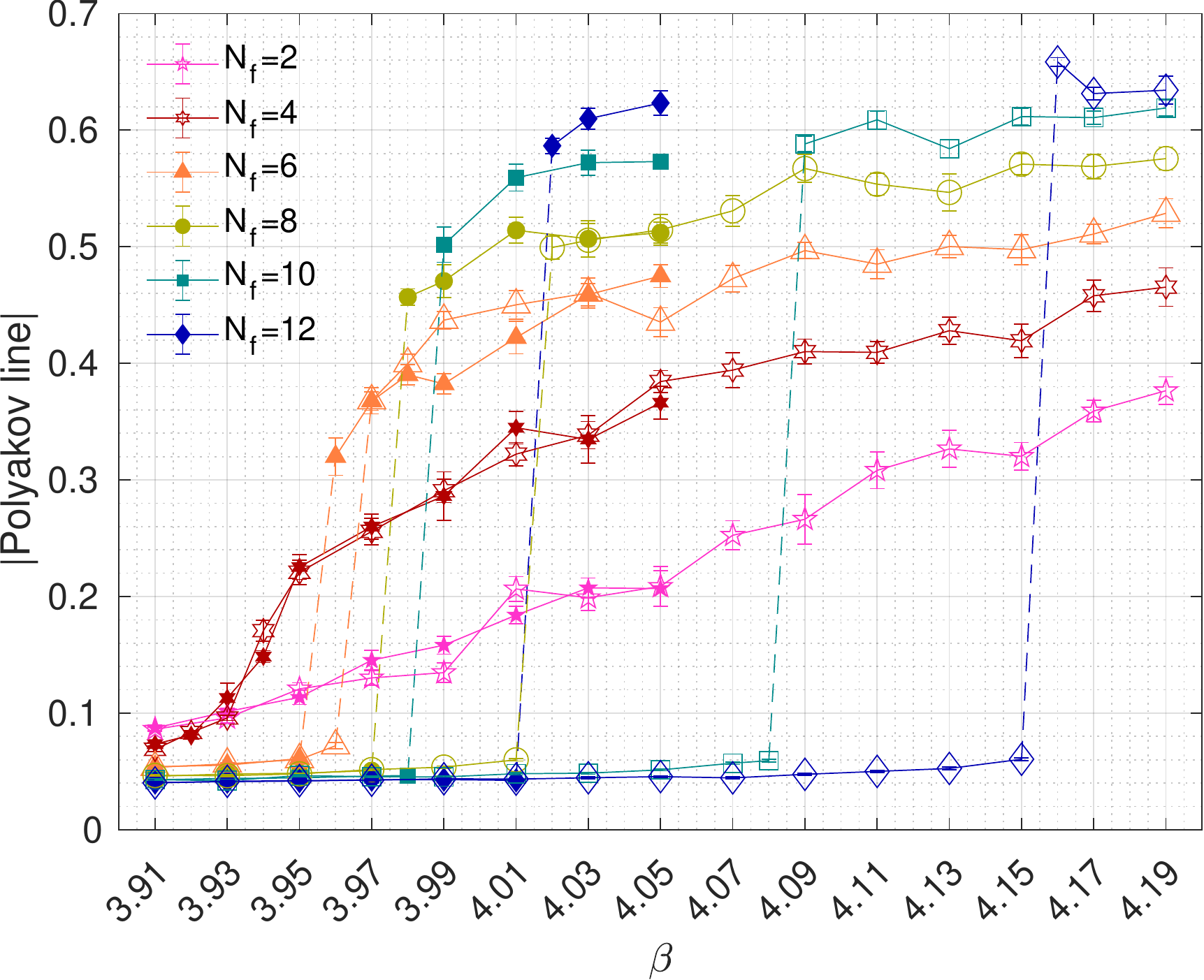}
  \caption{Comparison of the absolute value of the Polyakov line on $8^4$ volumes as the function of the bare coupling for $N_f = 2 - 12$ flavors. By performing starts from weak and strong coupling we are resolving a hysteresis for $N_f\ge 6$. Both the discontinuity and the hysteresis width grows with increasing flavor number.}
    \label{Fig.Poly}
\end{figure}

\begin{figure}[tb]
  \includegraphics[width=\columnwidth]{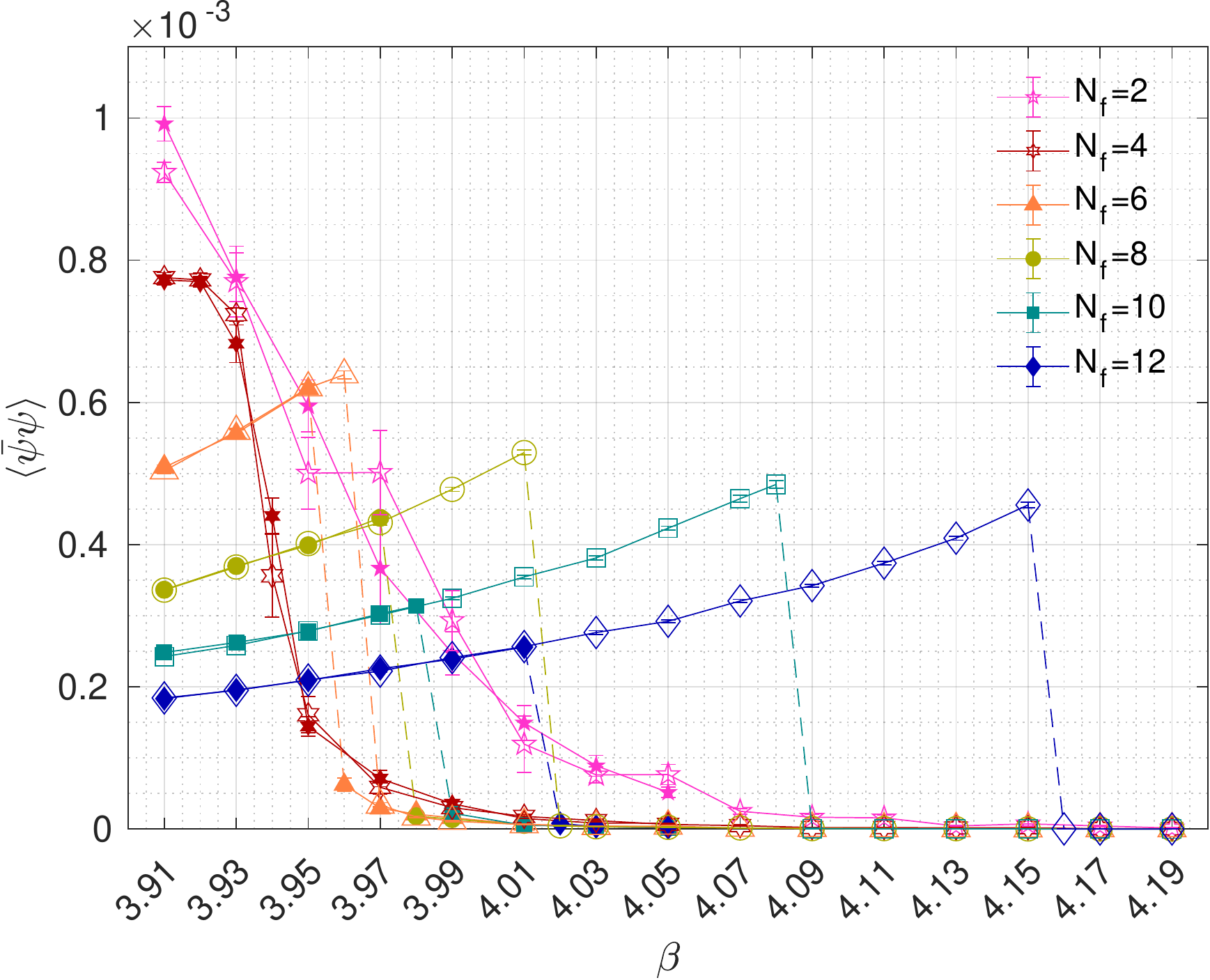}
  \caption{Comparison of the absolute value of the chiral condensate $\langle \bar \psi\psi\rangle$ on $8^4$ volumes as the function of the bare coupling for $N_f = 2 - 12$ flavors. By performing starts from weak and strong coupling we are resolving a hysteresis for $N_f\ge 6$. Both the discontinuity and the hysteresis width grows with increasing flavor number.}    
    \label{Fig.PBP}
\end{figure}

 We investigate the behavior of the plaquette, the Polyakov  line, the chiral condensate, and the finite volume renormalized gauge coupling in the $c=0.30$ scheme as the function of the bare coupling $\beta$. We have already discussed the renormalized gauge coupling in the Introduction, where in Fig.~\ref{Fig.g2} we show only the weak coupling branch. In Figs.~\ref{Fig.Plaquette}, \ref{Fig.Poly}, and \ref{Fig.PBP} we show the plaquette, the absolute value of the Polyakov line, and the chiral condensate both from the weak and strong coupling start simulations. All quantities show the bulk phase transition at identical bare couplings for $N_f\ge 6$, while $N_f=2$ and 4 are consistent with a cross over transition. The increased  width of the hysteresis loop is consistent with the increasing discontinuity of the phase transition for $N_f\ge 6$.
 
 The $N_f=2$ and 4 systems do not show any discontinuity, though at strong coupling both the Polyakov  line (Fig. \ref{Fig.Poly}) and the chiral condensate (Fig.~\ref{Fig.PBP}) indicate a transition from the deconfined weak to the confining strong coupling regime. This transition occurs at strong bare coupling where we expect the residual mass to be large, $am_\text{res}\gtrsim 0.1$. These simulations probe the system at finite mass and are not necessarily indicative of the finite temperature chiral transition.
We observe a very different behavior for $N_f\ge 6$. All observables indicate a first order phase transition from the deconfined phase with large Polyakov  line to a confined phase where the Polyakov  line is small (Fig.~\ref{Fig.Poly}). The chiral condensate also shows a transition from a chirally symmetric to a chirally broken regime, but the condensate is very different from the behavior observed for the $N_f=2$ and 4 flavor systems. After a discontinuity, $\langle \bar \psi \psi \rangle$  decreases as the gauge coupling gets stronger, while with small number of flavors we observe the opposite trend. At this point we cannot tell if we observe a new phase, possibly the analogue of the single site shift symmetry (S4) broken phase observed in many staggered fermion simulations \cite{Cheng:2011ic}, or the breakdown of the MDWF action where the mobility edge of the domain-wall kernel is comparable or below the domain wall height \cite{Golterman:2004cy,Golterman:2005fe}. Simulations with improved actions where the first order phase transition occurs on smoother gauge configurations could clarify this uncertainty in the future. 

Independent on the nature of the bulk phase transition, it limits the accessible parameter range of the simulations. The finite volume gradient flow coupling is defined at a fixed fraction of the lattice volume, $\sqrt{8t}= c L$. Larger volumes allow larger flow times, thus larger renormalized couplings. In practice the smallest lattice volumes used in the analysis determines the strongest renormalized gauge coupling of the step scaling function. In Fig.~\ref{Fig.g2} we show $g^2_{c=0.3}$ on $8^4$ volumes. On larger volumes the gauge coupling at fixed $c$ increases, but its value is still limited. In addition, numerical simulations very close to the bulk transition could pick up scaling behavior characteristic to that transition. In this work we limited the bare couplings to $\beta \ge 4.02$  to avoid contamination from the bulk transition.

\section{Summary}
\label{Sec.Summary}
\begin{figure}[t]
  \includegraphics[width=0.98\columnwidth]{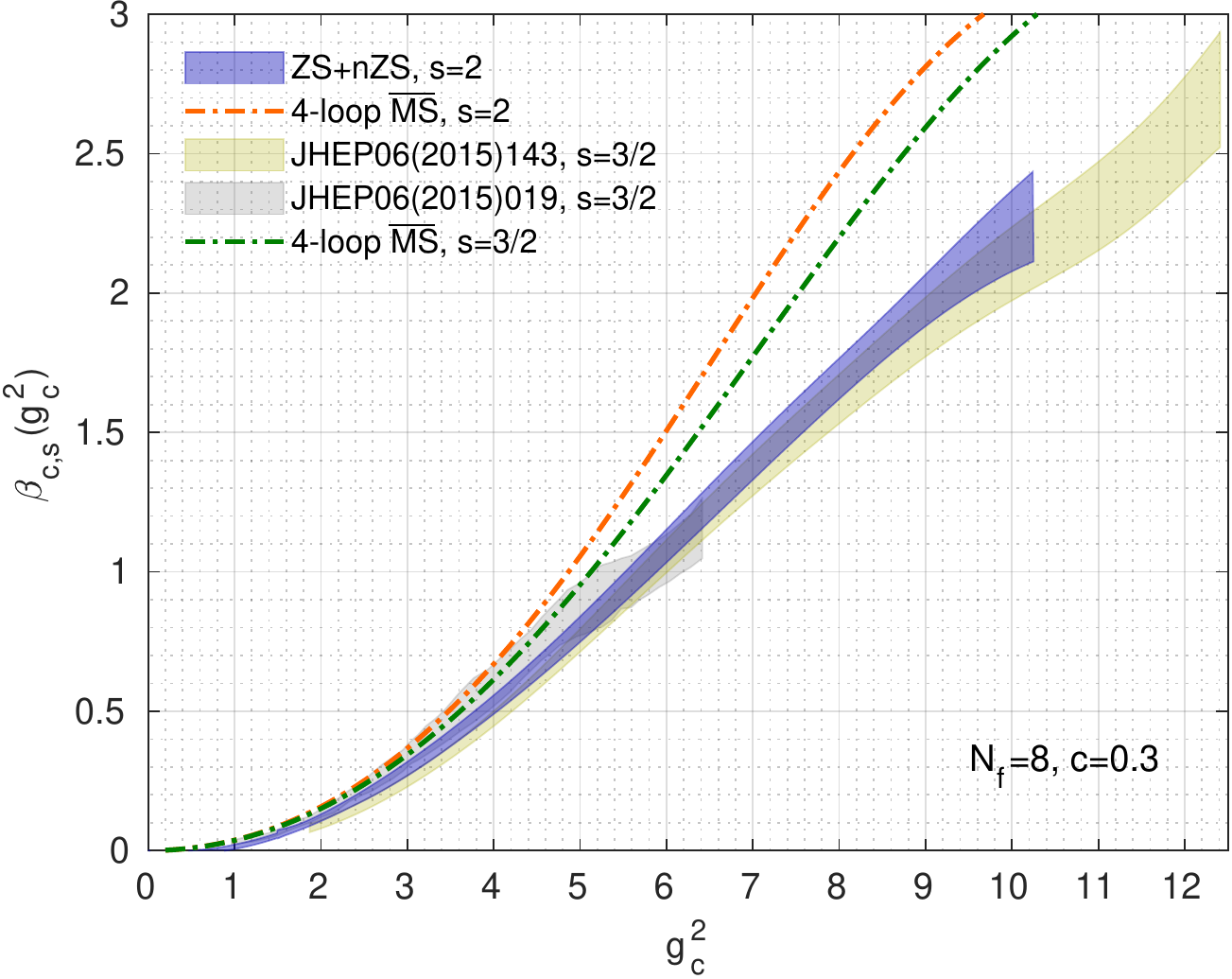}
  \caption{Comparison of our $N_f=8$ continuum results obtained from our preferred (n)ZS data set for $c = 0.300$ to the nonperturbative results obtained by Hasenfratz et al.~\cite{Hasenfratz:2014rna} and the Lattice Higgs Collaboration (LatHC) \cite{Fodor:2015baa}. While our results are based on the scale change $s=2$, both prior studies used $s=3/2$. }
  \label{Fig.Nf8cmp}
\end{figure}

In this work we have reported our results of the step scaling function of the SU(3) gauge $N_f=8$ fundamental flavors system. Our continuum limit results are consistent with prior calculations based on staggered lattice fermions  as shown in Fig.~\ref{Fig.Nf8cmp}. While all numerical results in Fig.~\ref{Fig.Nf8cmp} use the same gradient flow renormalization scheme $c=0.30$, the scale change is different. Judging from the differences between the 4-loop perturbative results in the $\overline{\text{MS}}$ scheme for $s=2$  and $s=1.5$ shown in Fig.~\ref{Fig.Nf8cmp}, we infer that the difference caused by switching from $s=2$ to $s=1.5$ in the nonperturbative numerical calculation could be of similar magnitude as the observed changes of the nonperturbative results. We tried to confirm this by repeating our analysis with $s=1.5$ forming the volume pairs $(8\to 12)$, $(16\to 24)$, and $(32\to 48)$ but unfortunately were not able to obtain a conclusive result. The two larger volume pairs, $(16\to 24)$ and $(32\to 48)$ turn out to be too noisy, whereas $8^4$ volumes are too small to be reliable in a linear continuum extrapolation in $a^2/L^2$.

This completes our first approach to investigate the renormalization group properties of SU(3) gauge systems with $N_f=2 - 12$ fundamental fermions using chirally symmetric M\"obius domain wall fermions and Symanzik gauge action. 

Most existing numerical simulations of many-flavor systems encounter a first order bulk phase transition at strong coupling.  This phase transition limits the parameter range of the simulations and restricts the strongest renormalized gauge coupling that can be reached at energy scales comparable to the inverse   lattice size. Comparison of results obtained using different lattice actions shows that the discontinuity of the bulk transition  depends strongly on the action. This suggests that, at least to some extent, the bulk phase transition is caused by strong ultraviolet lattice fluctuations, and improved lattice actions  may open up the parameter space allowing  to study  many-flavor systems at stronger gauge couplings.

The results presented in this work  reach up to $g_c^2\sim 10$, much below the possible continuous phase transition suggested in Ref.~\cite{Hasenfratz:2022qan}.  The phase diagram shows that the simulations are limited by a bulk first order transition and an improvement similar to the case of staggered fermions could help opening up the parameter space. This is, however, beyond the scope of the present work.

\begin{acknowledgments}

  We are very grateful to Peter Boyle, Guido Cossu, Anontin Portelli, and Azusa Yamaguchi who develop the \texttt{GRID} software library providing the basis of this work and who assisted us in installing and running \texttt{GRID} on different architectures and computing centers. A.H.~acknowledges support by DOE grant No.~DE-SC0010005 and C.R.~by DOE Grant No.~DE-SC0015845.

  Computations for this work were carried out in part on facilities of the USQCD Collaboration, which are funded by the Office of Science of the U.S.~Department of Energy, the RMACC Summit supercomputer \cite{UCsummit}, which is supported by the National Science Foundation (awards No.~ACI-1532235 and No.~ACI-1532236), the University of Colorado Boulder, and Colorado State University, and the \texttt{OMNI} cluster of the University of Siegen. This work used the Extreme Science and Engineering Discovery Environment (XSEDE), which is supported by National Science Foundation grant number ACI-1548562 \cite{xsede} through allocation TG-PHY180005 on the XSEDE resource \texttt{stampede2}.  This research also used resources of the National Energy Research Scientific Computing Center (NERSC), a U.S. Department of Energy Office of Science User Facility operated under Contract  No. DE-AC02-05CH11231. This document was prepared using the resources of the USQCD Collaboration at the Fermi National Accelerator Laboratory (Fermilab), a U.S. Department of Energy, Office of Science, HEP User Facility. Fermilab is managed by Fermi Research Alliance, LLC (FRA), acting under Contract No.~DE-AC02-07CH11359. We thank  Brookhaven National Laboratory (BNL), Fermilab,  Jefferson Lab, NERSC, the University of Colorado Boulder, the University of Siegen, TACC, the NSF, and the U.S.~DOE for providing the facilities essential for the completion of this work.
\end{acknowledgments}

\appendix
\setlength{\LTcapwidth}{\textwidth}
\section{\texorpdfstring{Renormalized couplings $g_c^2$}{Renormalized couplings gc2} and details of the polynomial interpolation}
\label{Sec.RenCouplings}

\begin{longtable*}{cccccccccccc}
  \caption{Details of our preferred analysis for $N_f=8$ based on Zeuthen flow and Symanzik operator. For each ensemble specified by the spatial extent $L/a$ and bare gauge coupling $\beta$ we list the number of measurements $N$ as well as the renormalized couplings $g_c^2$ for the analysis with (nZS) and without tree-level improvement (ZS) for the three renormalization schemes $c=0.300$, 0.275 and 0.250. In addition the integrated autocorrelation times estimated using the $\Gamma$-method \cite{Wolff:2003sm} are listed in units of 10 MDTU.}
  \label{Tab.Nf8_nZS_ZS}\\
  
  \hline\hline
      &         &     & \multicolumn{3}{c}{$c=0.300$}&\multicolumn{3}{c}{$c=0.275$}&\multicolumn{3}{c}{$c=0.250$}\\
  $L/a$ & $\beta$ & $N$ & $g_c^2$(nZS) & $g_c^2$(ZS) & $\tau_\text{int}$& $g_c^2$(nZS) &  $g_c^2$(ZS)  & $\tau_\text{int}$ &$g_c^2$(nZS)  &$g_c^2$(ZS)  & $\tau_\text{int}$\\
  \hline
  \endfirsthead

  \hline
      &         &     & \multicolumn{3}{c}{$c=0.300$}&\multicolumn{3}{c}{$c=0.275$}&\multicolumn{3}{c}{$c=0.250$}\\
  $L/a$ & $\beta$ & $N$ & $g_c^2$(nZS) &  $g_c^2$(ZS) & $\tau_\text{int}$& $g_c^2$(nZS) &$g_c^2$(ZS) & $\tau_\text{int}$ & $g_c^2$(nZS) & $g_c^2$(ZS) & $\tau_\text{int}$\\
  \hline
  \endhead

  \hline
  \endfoot

  \hline \hline
  \endlastfoot
  8 & 7.00 & 991 & 1.4494(21)  & 1.5233(22) & 0.51(5) & 1.4427(17)  & 1.5506(18) & 0.50(5) & 1.4342(14)  & 1.5962(15) & 0.50(5)\\ 
 8 & 6.50 & 1041 & 1.6582(24)  & 1.7427(25) & 0.53(6) & 1.6486(20)  & 1.7719(21) & 0.52(6) & 1.6368(16)  & 1.8217(17) & 0.52(6)\\ 
 8 & 6.00 & 1001 & 1.9504(28)  & 2.0498(30) & 0.51(6) & 1.9354(24)  & 2.0801(26) & 0.54(6) & 1.9173(19)  & 2.1339(21) & 0.53(6)\\ 
 8 & 5.50 & 1041 & 2.3478(35)  & 2.4674(36) & 0.53(7) & 2.3271(29)  & 2.5011(31) & 0.56(6) & 2.3020(24)  & 2.5620(26) & 0.56(6)\\ 
 8 & 5.00 & 1091 & 2.9989(48)  & 3.1517(50) & 0.61(7) & 2.9627(37)  & 3.1842(40) & 0.54(6) & 2.9196(29)  & 3.2494(32) & 0.50(4)\\ 
 8 & 4.70 & 1091 & 3.6321(64)  & 3.8172(67) & 0.62(8) & 3.5762(52)  & 3.8436(56) & 0.61(8) & 3.5107(42)  & 3.9072(47) & 0.60(8)\\ 
 8 & 4.50 & 1001 & 4.2508(82)  & 4.4675(86) & 0.68(9) & 4.1754(67)  & 4.4877(73) & 0.66(9) & 4.0861(55)  & 4.5477(61) & 0.65(9)\\ 
 8 & 4.40 & 862 & 4.6547(93)  & 4.8919(98) & 0.62(8) & 4.5647(76)  & 4.9060(82) & 0.59(8) & 4.4583(61)  & 4.9619(67) & 0.56(7)\\ 
 8 & 4.30 & 1031 & 5.2072(98)  & 5.473(10) & 0.62(8) & 5.0962(81)  & 5.4773(87) & 0.60(7) & 4.9627(66)  & 5.5232(73) & 0.58(7)\\ 
 8 & 4.25 & 957 & 5.560(13)  & 5.843(14) & 0.7(1) & 5.432(11)  & 5.838(12) & 0.7(1) & 5.2778(93)  & 5.874(10) & 0.7(1)\\ 
 8 & 4.20 & 637 & 5.985(16)  & 6.290(17) & 0.57(9) & 5.840(13)  & 6.277(14) & 0.56(8) & 5.661(11)  & 6.301(12) & 0.55(7)\\ 
 8 & 4.15 & 415 & 6.555(23)  & 6.889(24) & 0.51(7) & 6.373(21)  & 6.849(22) & 0.57(10) & 6.149(17)  & 6.844(19) & 0.58(10)\\ 
 8 & 4.10 & 405 & 7.390(37)  & 7.766(39) & 0.9(2) & 7.144(32)  & 7.678(34) & 0.9(2) & 6.840(27)  & 7.612(30) & 1.0(2)\\ 
 8 & 4.05 & 398 & 8.741(63)  & 9.186(66) & 1.2(3) & 8.365(54)  & 8.990(58) & 1.3(3) & 7.903(45)  & 8.796(50) & 1.4(4)\\ 
 8 & 4.03 & 368 & 9.669(95)  & 10.16(10) & 1.6(5) & 9.189(80)  & 9.876(86) & 1.5(5) & 8.617(68)  & 9.591(75) & 1.6(5)\\ 
 8 & 4.02 & 356 & 10.06(12)  & 10.57(13) & 1.8(6) & 9.56(11)  & 10.27(11) & 1.9(6) & 8.940(85)  & 9.950(94) & 1.8(6)\\ 
 \hline 
10 & 7.00 & 605 & 1.4735(28)  & 1.5026(29) & 0.58(9) & 1.4663(24)  & 1.5077(24) & 0.59(9) & 1.4576(20)  & 1.5193(21) & 0.59(10)\\ 
 10 & 6.50 & 605 & 1.6971(36)  & 1.7307(37) & 0.7(1) & 1.6862(30)  & 1.7339(31) & 0.6(1) & 1.6735(24)  & 1.7443(25) & 0.62(10)\\ 
 10 & 6.00 & 605 & 1.9964(41)  & 2.0358(42) & 0.57(8) & 1.9801(32)  & 2.0361(33) & 0.50(4) & 1.9619(26)  & 2.0449(27) & 0.49(4)\\ 
 10 & 5.50 & 605 & 2.4311(49)  & 2.4791(50) & 0.55(8) & 2.4060(39)  & 2.4740(40) & 0.51(6) & 2.3781(31)  & 2.4788(32) & 0.49(6)\\ 
 10 & 5.00 & 605 & 3.1295(69)  & 3.1913(70) & 0.61(10) & 3.0885(56)  & 3.1757(57) & 0.57(8) & 3.0428(44)  & 3.1716(45) & 0.54(8)\\ 
 10 & 4.70 & 605 & 3.7937(74)  & 3.8686(75) & 0.51(6) & 3.7352(59)  & 3.8407(61) & 0.47(5) & 3.6707(49)  & 3.8261(51) & 0.49(4)\\ 
 10 & 4.50 & 605 & 4.4778(91)  & 4.5662(93) & 0.51(7) & 4.3981(75)  & 4.5223(77) & 0.49(7) & 4.3103(61)  & 4.4927(63) & 0.49(7)\\ 
 10 & 4.40 & 605 & 4.962(12)  & 5.060(13) & 0.7(1) & 4.8611(97)  & 4.9985(100) & 0.59(9) & 4.7516(75)  & 4.9528(78) & 0.53(9)\\ 
 10 & 4.30 & 605 & 5.536(17)  & 5.645(17) & 1.0(2) & 5.422(14)  & 5.575(15) & 1.0(2) & 5.296(11)  & 5.520(12) & 0.9(2)\\ 
 10 & 4.20 & 605 & 6.415(17)  & 6.542(18) & 0.7(1) & 6.268(14)  & 6.446(15) & 0.6(1) & 6.106(12)  & 6.365(12) & 0.62(10)\\ 
 10 & 4.15 & 603 & 7.045(20)  & 7.185(21) & 0.8(1) & 6.881(17)  & 7.075(18) & 0.8(1) & 6.691(14)  & 6.974(15) & 0.7(1)\\ 
 10 & 4.10 & 600 & 7.917(39)  & 8.073(40) & 1.8(5) & 7.727(33)  & 7.945(34) & 1.7(4) & 7.498(27)  & 7.815(28) & 1.5(4)\\ 
 10 & 4.05 & 584 & 9.443(53)  & 9.629(54) & 1.6(4) & 9.202(46)  & 9.462(48) & 1.4(3) & 8.868(38)  & 9.243(40) & 1.3(3)\\ 
 10 & 4.03 & 567 & 10.596(54)  & 10.806(56) & 1.1(3) & 10.281(50)  & 10.572(51) & 1.2(3) & 9.863(45)  & 10.281(47) & 1.1(3)\\ 
 \hline 
12 & 7.00 & 487 & 1.5027(36)  & 1.5170(36) & 0.7(1) & 1.4937(28)  & 1.5137(28) & 0.7(1) & 1.4832(21)  & 1.5123(22) & 0.6(1)\\ 
 12 & 6.50 & 498 & 1.7292(40)  & 1.7456(40) & 0.7(1) & 1.7171(31)  & 1.7402(32) & 0.6(1) & 1.7031(24)  & 1.7365(25) & 0.53(8)\\ 
 12 & 6.00 & 495 & 2.0341(48)  & 2.0535(48) & 0.7(1) & 2.0179(41)  & 2.0449(41) & 0.7(1) & 1.9991(33)  & 2.0383(34) & 0.7(1)\\ 
 12 & 5.50 & 491 & 2.4939(59)  & 2.5176(59) & 0.6(1) & 2.4676(46)  & 2.5007(46) & 0.57(9) & 2.4383(36)  & 2.4862(37) & 0.52(8)\\ 
 12 & 5.00 & 491 & 3.224(11)  & 3.255(11) & 1.1(3) & 3.1835(82)  & 3.2262(83) & 0.9(2) & 3.1376(60)  & 3.1992(62) & 0.8(2)\\ 
 12 & 4.70 & 494 & 3.941(12)  & 3.978(13) & 1.1(2) & 3.879(10)  & 3.931(11) & 1.1(2) & 3.8115(85)  & 3.8863(87) & 1.1(2)\\ 
 12 & 4.40 & 466 & 5.162(21)  & 5.211(21) & 1.3(3) & 5.063(16)  & 5.131(16) & 1.1(3) & 4.956(13)  & 5.053(13) & 1.0(2)\\ 
 12 & 4.30 & 467 & 5.859(23)  & 5.914(24) & 1.1(3) & 5.725(18)  & 5.802(18) & 1.0(2) & 5.584(14)  & 5.693(15) & 0.9(2)\\ 
 12 & 4.20 & 491 & 6.798(22)  & 6.862(22) & 0.9(2) & 6.638(18)  & 6.727(18) & 0.8(2) & 6.469(14)  & 6.596(15) & 0.8(2)\\ 
 12 & 4.15 & 490 & 7.429(32)  & 7.500(33) & 1.3(3) & 7.261(27)  & 7.358(27) & 1.2(3) & 7.081(22)  & 7.220(22) & 1.1(3)\\ 
 12 & 4.10 & 590 & 8.397(30)  & 8.477(30) & 1.0(2) & 8.205(24)  & 8.315(24) & 0.9(2) & 8.004(19)  & 8.161(19) & 0.8(1)\\ 
 12 & 4.05 & 577 & 9.866(48)  & 9.959(48) & 1.6(4) & 9.689(42)  & 9.819(42) & 1.6(4) & 9.468(35)  & 9.654(35) & 1.4(3)\\ 
 12 & 4.03 & 557 & 10.895(54)  & 10.999(54) & 1.5(4) & 10.711(52)  & 10.854(52) & 1.6(4) & 10.487(50)  & 10.693(51) & 1.6(4)\\ 
 \hline 
16 & 7.00 & 592 & 1.5479(34)  & 1.5526(34) & 0.7(1) & 1.5363(25)  & 1.5429(25) & 0.56(8) & 1.5238(20)  & 1.5333(20) & 0.52(7)\\ 
 16 & 6.50 & 555 & 1.7838(44)  & 1.7893(44) & 0.8(1) & 1.7696(35)  & 1.7772(35) & 0.7(1) & 1.7541(28)  & 1.7650(28) & 0.7(1)\\ 
 16 & 6.00 & 305 & 2.1350(89)  & 2.1415(90) & 1.2(3) & 2.1112(69)  & 2.1202(69) & 1.0(3) & 2.0857(51)  & 2.0987(52) & 0.8(2)\\ 
 16 & 5.50 & 195 & 2.6065(93)  & 2.6145(94) & 0.6(2) & 2.5759(78)  & 2.5870(79) & 0.6(2) & 2.5429(65)  & 2.5588(65) & 0.6(2)\\ 
 16 & 5.00 & 339 & 3.379(17)  & 3.390(17) & 1.6(5) & 3.335(14)  & 3.350(14) & 1.5(5) & 3.287(11)  & 3.307(11) & 1.4(4)\\ 
 16 & 4.70 & 431 & 4.221(18)  & 4.234(18) & 1.4(4) & 4.143(14)  & 4.161(14) & 1.2(3) & 4.062(11)  & 4.087(11) & 1.2(3)\\ 
 16 & 4.50 & 208 & 5.015(40)  & 5.030(40) & 2.4(10) & 4.910(31)  & 4.931(31) & 2.0(8) & 4.802(23)  & 4.832(23) & 1.7(6)\\ 
 16 & 4.40 & 261 & 5.578(47)  & 5.595(47) & 3(1) & 5.456(36)  & 5.480(37) & 3(1) & 5.330(27)  & 5.363(28) & 2.4(9)\\ 
 16 & 4.30 & 369 & 6.258(23)  & 6.277(23) & 1.0(3) & 6.119(18)  & 6.145(18) & 0.9(2) & 5.974(14)  & 6.012(14) & 0.8(2)\\ 
 16 & 4.25 & 232 & 6.728(46)  & 6.748(46) & 1.9(7) & 6.574(37)  & 6.602(37) & 1.9(7) & 6.413(31)  & 6.453(31) & 1.9(7)\\ 
 16 & 4.20 & 481 & 7.282(27)  & 7.304(27) & 1.3(3) & 7.106(21)  & 7.136(21) & 1.1(3) & 6.924(17)  & 6.967(17) & 1.0(2)\\ 
 16 & 4.15 & 487 & 8.082(34)  & 8.107(34) & 1.4(4) & 7.867(26)  & 7.901(26) & 1.3(3) & 7.650(20)  & 7.698(21) & 1.2(3)\\ 
 16 & 4.10 & 569 & 9.054(46)  & 9.082(46) & 2.5(7) & 8.817(36)  & 8.855(36) & 2.2(6) & 8.583(28)  & 8.637(28) & 2.0(5)\\ 
 16 & 4.05 & 443 & 10.623(38)  & 10.655(38) & 1.1(3) & 10.365(33)  & 10.409(34) & 1.2(3) & 10.121(29)  & 10.184(29) & 1.2(3)\\ 
 16 & 4.03 & 356 & 11.484(64)  & 11.520(64) & 1.6(5) & 11.239(56)  & 11.288(56) & 1.7(5) & 11.032(46)  & 11.100(46) & 1.5(5)\\ 
 16 & 4.02 & 515 & 12.139(54)  & 12.176(54) & 1.5(4) & 11.920(46)  & 11.971(47) & 1.4(3) & 11.736(42)  & 11.809(42) & 1.3(3)\\ 
 \hline 
20 & 7.00 & 291 & 1.5782(65)  & 1.5802(65) & 1.1(3) & 1.5665(52)  & 1.5693(52) & 1.1(3) & 1.5539(38)  & 1.5579(38) & 0.9(2)\\ 
 20 & 6.50 & 220 & 1.829(11)  & 1.831(11) & 1.7(6) & 1.8134(89)  & 1.8166(89) & 1.6(6) & 1.7966(69)  & 1.8013(69) & 1.4(5)\\ 
 20 & 6.00 & 165 & 2.154(11)  & 2.157(11) & 0.9(3) & 2.1360(92)  & 2.1398(92) & 0.9(3) & 2.1151(75)  & 2.1206(76) & 0.9(3)\\ 
 20 & 5.50 & 164 & 2.675(20)  & 2.678(20) & 2.1(9) & 2.644(15)  & 2.649(15) & 1.8(7) & 2.611(11)  & 2.617(11) & 1.5(6)\\ 
 20 & 5.00 & 128 & 3.554(32)  & 3.559(32) & 1.9(9) & 3.497(25)  & 3.504(25) & 1.6(7) & 3.437(18)  & 3.446(19) & 1.4(5)\\ 
 20 & 4.70 & 271 & 4.416(23)  & 4.422(23) & 1.2(4) & 4.328(18)  & 4.335(18) & 1.1(3) & 4.237(14)  & 4.248(14) & 1.1(3)\\ 
 20 & 4.50 & 271 & 5.288(28)  & 5.294(28) & 1.7(6) & 5.169(21)  & 5.178(21) & 1.5(5) & 5.048(17)  & 5.061(17) & 1.4(5)\\ 
 20 & 4.40 & 271 & 5.879(42)  & 5.887(43) & 2.3(9) & 5.746(31)  & 5.756(31) & 1.9(7) & 5.608(23)  & 5.623(23) & 1.6(6)\\ 
 20 & 4.30 & 271 & 6.703(60)  & 6.712(61) & 3(1) & 6.532(46)  & 6.543(46) & 3(1) & 6.357(32)  & 6.373(33) & 2.3(8)\\ 
 20 & 4.20 & 271 & 7.855(76)  & 7.864(76) & 4(2) & 7.630(56)  & 7.643(56) & 3(1) & 7.408(39)  & 7.427(39) & 3(1)\\ 
 20 & 4.15 & 257 & 8.627(50)  & 8.638(50) & 1.9(7) & 8.374(39)  & 8.389(39) & 1.7(6) & 8.124(30)  & 8.145(30) & 1.6(5)\\ 
 20 & 4.10 & 269 & 9.643(57)  & 9.655(57) & 1.8(6) & 9.359(45)  & 9.375(45) & 1.6(6) & 9.079(34)  & 9.102(34) & 1.4(5)\\ 
 20 & 4.05 & 267 & 11.23(11)  & 11.25(11) & 4(2) & 10.927(87)  & 10.946(87) & 4(2) & 10.626(61)  & 10.653(62) & 3(1)\\ 
 20 & 4.03 & 236 & 12.299(75)  & 12.315(75) & 1.9(7) & 11.931(59)  & 11.952(59) & 1.7(6) & 11.621(48)  & 11.651(48) & 1.5(5)\\ 
 \hline 
24 & 7.00 & 323 & 1.6101(81)  & 1.6111(81) & 2.0(6) & 1.5963(58)  & 1.5977(58) & 1.4(4) & 1.5816(42)  & 1.5836(43) & 1.2(3)\\ 
 24 & 6.50 & 315 & 1.866(11)  & 1.867(11) & 2.5(9) & 1.8483(82)  & 1.8500(82) & 2.0(6) & 1.8299(63)  & 1.8322(63) & 1.7(5)\\ 
 24 & 6.00 & 212 & 2.233(13)  & 2.234(13) & 1.6(6) & 2.208(11)  & 2.210(11) & 1.5(5) & 2.1811(85)  & 2.1838(86) & 1.4(5)\\ 
 24 & 5.50 & 261 & 2.805(19)  & 2.807(19) & 2.5(10) & 2.761(15)  & 2.763(15) & 2.1(8) & 2.715(11)  & 2.718(11) & 1.8(6)\\ 
 24 & 5.00 & 296 & 3.683(19)  & 3.685(19) & 1.7(5) & 3.617(14)  & 3.621(14) & 1.4(4) & 3.550(11)  & 3.554(11) & 1.3(4)\\ 
 24 & 4.70 & 259 & 4.603(25)  & 4.605(25) & 1.7(6) & 4.508(20)  & 4.512(20) & 1.6(6) & 4.410(16)  & 4.415(16) & 1.5(5)\\ 
 24 & 4.50 & 315 & 5.588(39)  & 5.591(39) & 3(1) & 5.452(30)  & 5.457(30) & 2.7(10) & 5.313(23)  & 5.320(23) & 2.4(9)\\ 
 24 & 4.40 & 208 & 6.177(40)  & 6.180(40) & 1.8(7) & 6.024(31)  & 6.029(31) & 1.6(6) & 5.868(24)  & 5.876(24) & 1.5(5)\\ 
 24 & 4.30 & 230 & 7.005(86)  & 7.009(86) & 4(2) & 6.827(65)  & 6.833(65) & 3(2) & 6.646(47)  & 6.654(47) & 3(1)\\ 
 24 & 4.20 & 281 & 8.258(59)  & 8.263(59) & 3(1) & 8.026(47)  & 8.033(47) & 3(1) & 7.790(36)  & 7.800(36) & 2.5(9)\\ 
 24 & 4.15 & 269 & 9.094(91)  & 9.100(91) & 5(2) & 8.819(69)  & 8.827(69) & 4(2) & 8.544(52)  & 8.554(52) & 4(2)\\ 
 24 & 4.10 & 282 & 10.157(94)  & 10.163(94) & 4(2) & 9.840(72)  & 9.849(72) & 4(1) & 9.533(55)  & 9.545(55) & 3(1)\\ 
 24 & 4.05 & 269 & 11.984(78)  & 11.991(78) & 1.9(7) & 11.570(60)  & 11.580(60) & 1.7(6) & 11.187(46)  & 11.201(46) & 1.5(5)\\ 
 24 & 4.03 & 255 & 12.917(94)  & 12.925(94) & 2.1(8) & 12.514(68)  & 12.525(68) & 1.7(6) & 12.139(55)  & 12.154(55) & 1.6(6)\\ 
 \hline 
32 & 7.00 & 216 & 1.657(13)  & 1.657(13) & 3(1) & 1.6413(100)  & 1.6417(100) & 3(1) & 1.6249(77)  & 1.6256(77) & 2.3(10)\\ 
 32 & 6.50 & 201 & 1.910(12)  & 1.910(12) & 2.0(8) & 1.8949(91)  & 1.8954(91) & 1.7(6) & 1.8777(67)  & 1.8785(67) & 1.3(5)\\ 
 32 & 6.00 & 201 & 2.326(14)  & 2.327(14) & 2.0(8) & 2.296(12)  & 2.296(12) & 1.8(7) & 2.2640(92)  & 2.2649(92) & 1.7(6)\\ 
 32 & 5.50 & 201 & 2.887(38)  & 2.888(38) & 6(3) & 2.851(31)  & 2.852(31) & 6(3) & 2.811(24)  & 2.812(24) & 5(3)\\ 
 32 & 5.00 & 203 & 3.922(43)  & 3.923(43) & 6(3) & 3.848(32)  & 3.849(32) & 4(2) & 3.771(23)  & 3.772(23) & 4(2)\\ 
 32 & 4.70 & 205 & 4.980(83)  & 4.981(83) & 9(5) & 4.850(64)  & 4.851(64) & 8(4) & 4.721(48)  & 4.723(48) & 7(4)\\ 
 32 & 4.40 & 201 & 6.80(11)  & 6.80(11) & 8(5) & 6.607(81)  & 6.608(81) & 7(4) & 6.408(56)  & 6.411(56) & 5(3)\\ 
 32 & 4.30 & 201 & 7.710(69)  & 7.711(69) & 3(1) & 7.488(56)  & 7.490(56) & 3(1) & 7.259(43)  & 7.262(44) & 3(1)\\ 
 32 & 4.20 & 332 & 9.204(67)  & 9.206(67) & 5(2) & 8.875(50)  & 8.877(50) & 4(2) & 8.555(36)  & 8.559(36) & 3(1)\\ 
 32 & 4.15 & 171 & 10.04(11)  & 10.04(11) & 3(1) & 9.689(86)  & 9.692(86) & 3(1) & 9.345(65)  & 9.348(65) & 2(1)\\ 
 32 & 4.10 & 329 & 11.45(14)  & 11.46(14) & 7(3) & 11.02(11)  & 11.02(11) & 6(3) & 10.595(79)  & 10.600(80) & 5(2)\\ 
 32 & 4.05 & 271 & 13.21(14)  & 13.21(14) & 6(3) & 12.689(98)  & 12.693(98) & 5(2) & 12.209(72)  & 12.214(73) & 4(2)\\ 
 \hline 
48 & 7.00 & 101 & 1.714(21)  & 1.714(21) & 6(3) & 1.702(16)  & 1.703(16) & 5(3) & 1.689(11)  & 1.689(11) & 3(2)\\ 
 48 & 6.50 & 100 & 2.008(13)  & 2.009(13) & 1.6(8) & 1.990(12)  & 1.991(12) & 1.8(9) & 1.970(11)  & 1.971(11) & 1.9(10)\\ 
 48 & 6.00 & 101 & 2.486(32)  & 2.486(32) & 5(3) & 2.451(27)  & 2.451(27) & 5(3) & 2.413(23)  & 2.414(23) & 5(3)\\ 
 48 & 5.50 & 100 & 3.107(61)  & 3.107(61) & 6(3) & 3.059(55)  & 3.059(55) & 6(3) & 3.008(49)  & 3.009(49) & 6(3)\\ 
 48 & 5.00 & 100 & 4.326(86)  & 4.326(86) & 7(4) & 4.239(67)  & 4.239(67) & 7(4) & 4.144(49)  & 4.144(49) & 5(3)\\ 
 48 & 4.70 & 102 & 5.54(15)  & 5.55(15) & 9(4) & 5.39(11)  & 5.39(11) & 9(4) & 5.237(79)  & 5.237(79) & 8(4)\\ 
 48 & 4.40 & 131 & 7.81(13)  & 7.81(13) & 6(3) & 7.534(95)  & 7.534(95) & 6(3) & 7.266(70)  & 7.267(70) & 5(3)\\ 
 48 & 4.30 & 91 & 8.535(97)  & 8.536(97) & 4(2) & 8.321(80)  & 8.321(80) & 3(2) & 8.090(62)  & 8.091(62) & 3(2)\\ 
 48 & 4.20 & 91 & 10.37(18)  & 10.37(18) & 6(3) & 10.04(14)  & 10.04(14) & 6(3) & 9.70(11)  & 9.71(11) & 5(3)\\ 
 48 & 4.15 & 87 & 11.63(16)  & 11.63(16) & 5(3) & 11.18(13)  & 11.18(13) & 5(3) & 10.744(99)  & 10.745(99) & 4(2)\\ 
 48 & 4.10 & 96 & 13.00(26)  & 13.00(26) & 7(4) & 12.54(19)  & 12.54(19) & 7(4) & 12.08(14)  & 12.09(14) & 6(3)

\end{longtable*}

\begin{minipage}{\textwidth}
  \captionof{table}{Results of the interpolation fits for the five $N_f=8$ lattice volume pairs for our preferred ZS (top half) and nZS (bottom half) analysis using renormalization schemes $c = 0.300$, $0.275$, and $0.250$. Since discretization effects are sufficiently small for nZS, we constrain the constant term $b_0 = 0$ in Eq.~(\ref{Eq.fit_form}) whereas for ZS the intercept $b_0$ is fitted. In addition we list the degree of freedom (d.o.f.), $\chi^2/\text{d.o.f.}$~as well as the $p$-value.}
 \label{Tab.interpolationsNf8}
  \begin{tabular}{c@{~~~~}ccc@{~~}c@{~~}ccccc}
    \hline \hline
    & analysis &    $c$ &d.o.f. &$\chi^2$/d.o.f.&$p$-value&    $b_3$        & $b_2$     & $b_1$      & $b_0$ \\ \hline
$ 8\to 16$ & ZS & 0.300 & 12 & 0.764 & 0.689 & -0.00313(34) & 0.0489(50) &-0.071(21) &0.027(23) \\
$10\to 20$ & ZS & 0.300 & 10 & 0.431 & 0.932 & -0.00420(44) & 0.0637(69) &-0.107(30) &0.086(34) \\
$12\to 24$ & ZS & 0.300 &  9 & 0.386 & 0.943 & -0.00221(47) & 0.0347(74) &0.018(32) &-0.034(37) \\
$16\to 32$ & ZS & 0.300 &  8 & 0.608 & 0.772 & -0.00327(71) & 0.057(11)  &-0.064(50) &0.045(60) \\
$24\to 48$ & ZS & 0.300 &  7 & 0.960 & 0.459 & -0.0011(16)  & 0.025(23)  &0.07(10) &-0.10(12) \\
\hline
$ 8\to 16$ & ZS & 0.275 & 12 & 0.751 & 0.702 & -0.00237(31) & 0.0418(44) &-0.071(18) &0.013(19) \\
$10\to 20$ & ZS & 0.275 & 10 & 0.421 & 0.937 & -0.00407(38) & 0.0609(58) &-0.104(25) &0.076(28) \\
$12\to 24$ & ZS & 0.275 &  9 & 0.367 & 0.951 & -0.00258(38) & 0.0385(60) &-0.001(25) &-0.018(29) \\
$16\to 32$ & ZS & 0.275 &  8 & 0.609 & 0.771 & -0.00344(58) & 0.0569(90) &-0.066(40) &0.046(47) \\
$24\to 48$ & ZS & 0.275 &  7 & 0.931 & 0.481 & -0.0013(13)  & 0.027(20)  & 0.061(84) &-0.090(97)\\
\hline
$ 8\to 16$ & ZS & 0.250 & 12 & 0.774 & 0.679 &-0.00107(29) &0.0300(39) &-0.067(15) &-0.012(17) \\
$10\to 20$ & ZS & 0.250 & 10 & 0.441 & 0.927 &-0.00370(33) &0.0565(49) &-0.102(20) &0.065(22) \\
$12\to 24$ & ZS & 0.250 &  9 & 0.474 & 0.893 &-0.00282(32) &0.0410(49) &-0.017(20) &-0.007(23) \\
$16\to 32$ & ZS & 0.250 &  8 & 0.652 & 9.734 &-0.00361(48) &0.0571(72) &-0.068(32) &0.047(37) \\
$24\to 48$ & ZS & 0.250 &  7 & 0.932 & 0.480 &-0.0014(11)  &0.029(16)  &0.050(68)  &-0.074(77)\\
\hline \hline
$ 8\to 16$ & nZS & 0.300 & 13 & 0.835 & 0.623 &=-0.00321(19) &0.0478(18) &-0.0147(32) &---\\
$10\to 20$ & nZS & 0.300 & 11 & 0.999 & 0.445 &=-0.00338(20) &0.0488(23) &-0.0190(49) &---\\
$12\to 24$ & nZS & 0.300 & 10 & 0.433 & 0.931 &-0.00267(22)  &0.0419(26) &-0.0043(57) &---\\
$16\to 32$ & nZS & 0.300 &  9 & 0.604 & 0.795 &-0.00283(37)  &0.0490(40) &-0.0247(79) &---\\
$24\to 48$ & nZS & 0.300 &  8 & 0.935 & 0.486 &-0.00232(74)  &0.0445(79) &-0.017(15)  &---\\
\hline
$ 8\to 16$ & nZS & 0.275 & 13 & 0.763 & 0.701 &-0.00271(18)  &0.0450(16) &-0.0130(27) &---\\
$10\to 20$ & nZS & 0.275 & 11 & 1.085 & 0.369 &-0.00340(18)  &0.0483(19) &-0.0190(40) &---\\
$12\to 24$ & nZS & 0.275 & 10 & 0.371 & 0.960 &-0.00290(18)  &0.0431(21) &-0.0081(44) &---\\
$16\to 32$ & nZS & 0.275 &  9 & 0.647 & 0.757 &-0.00299(30)  &0.0491(32) &-0.0245(63) &---\\
$24\to 48$ & nZS & 0.275 &  8 & 0.923 & 0.496 &-0.00238(63)  &0.0444(65) &-0.016(12)  &---\\
\hline
$ 8\to 16$ & nZS & 0.250 & 13 & 0.813 & 0.647 &-0.00169(18)  &0.0400(15) &-0.0086(24) &---\\ 
$10\to 20$ & nZS & 0.250 & 11 & 1.163 & 0.307 &-0.00322(16)  &0.0468(16) &-0.0175(33) &---\\
$12\to 24$ & nZS & 0.250 & 10 & 0.439 & 0.928 &-0.00308(16)  &0.0441(17) &-0.0108(35) &---\\
$16\to 32$ & nZS & 0.250 &  9 & 0.758 & 0.655 &-0.00314(24)  &0.0491(25) &-0.0243(49) &---\\
$24\to 48$ & nZS & 0.250 &  8 & 0.931 & 0.489 &-0.00239(51)  &0.0437(52) &-0.0139(94) &---\\
\hline \hline
  \end{tabular}
\end{minipage}
\clearpage
\section{Analysis for \texorpdfstring{$c=0.275$ and 0.250}{c=0.275 and 0.250}}
\label{Sec.c0275_c0250}
\hspace*{-0.525\textwidth}\begin{minipage}{\textwidth}
  \begin{minipage}{0.49\textwidth}
   \flushright 
   \includegraphics[width=0.96\textwidth]{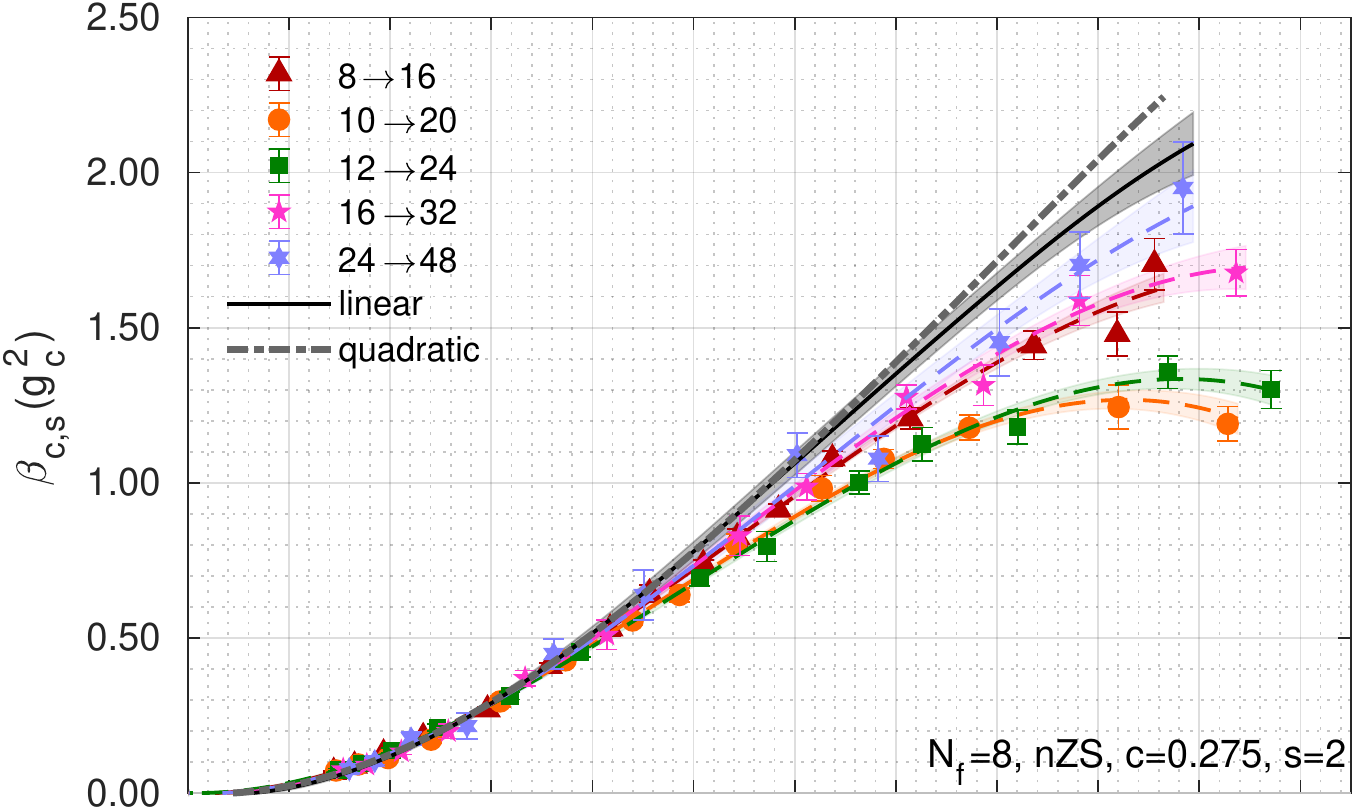}\\
   \includegraphics[width=0.932\textwidth]{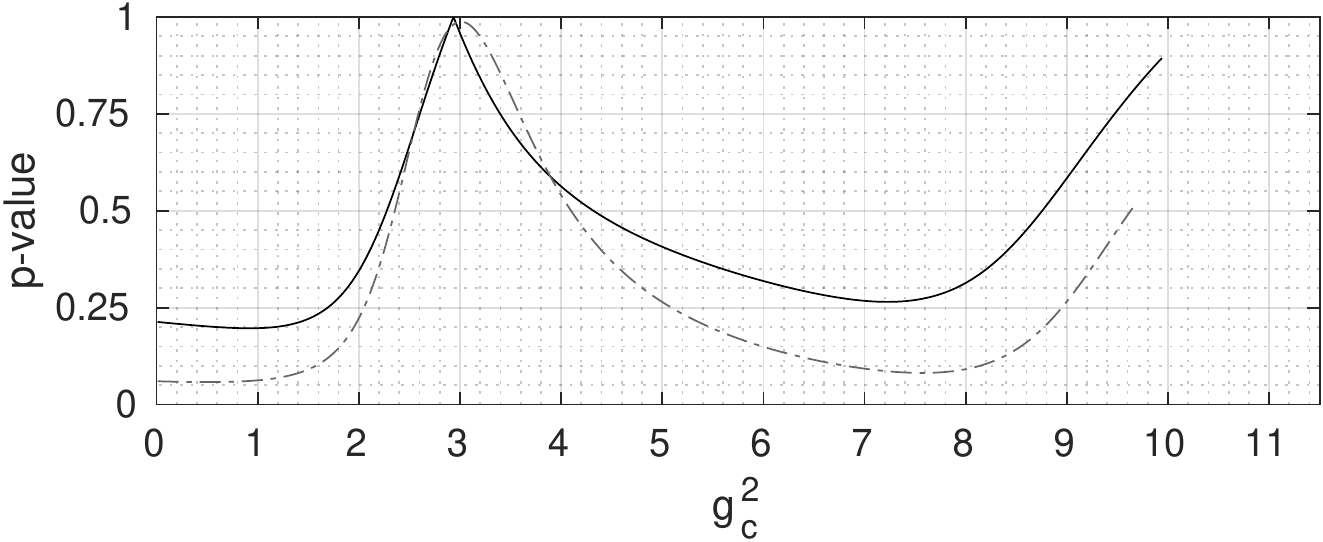} 
   \includegraphics[width=0.96\textwidth]{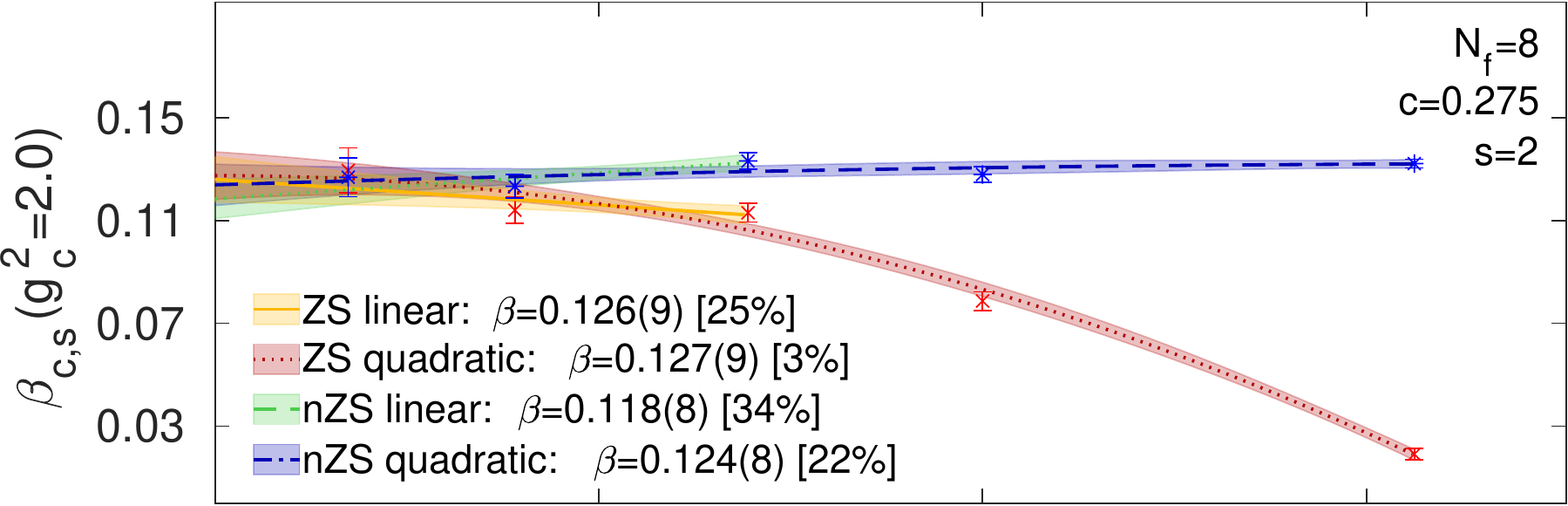}\\
   \includegraphics[width=0.96\textwidth]{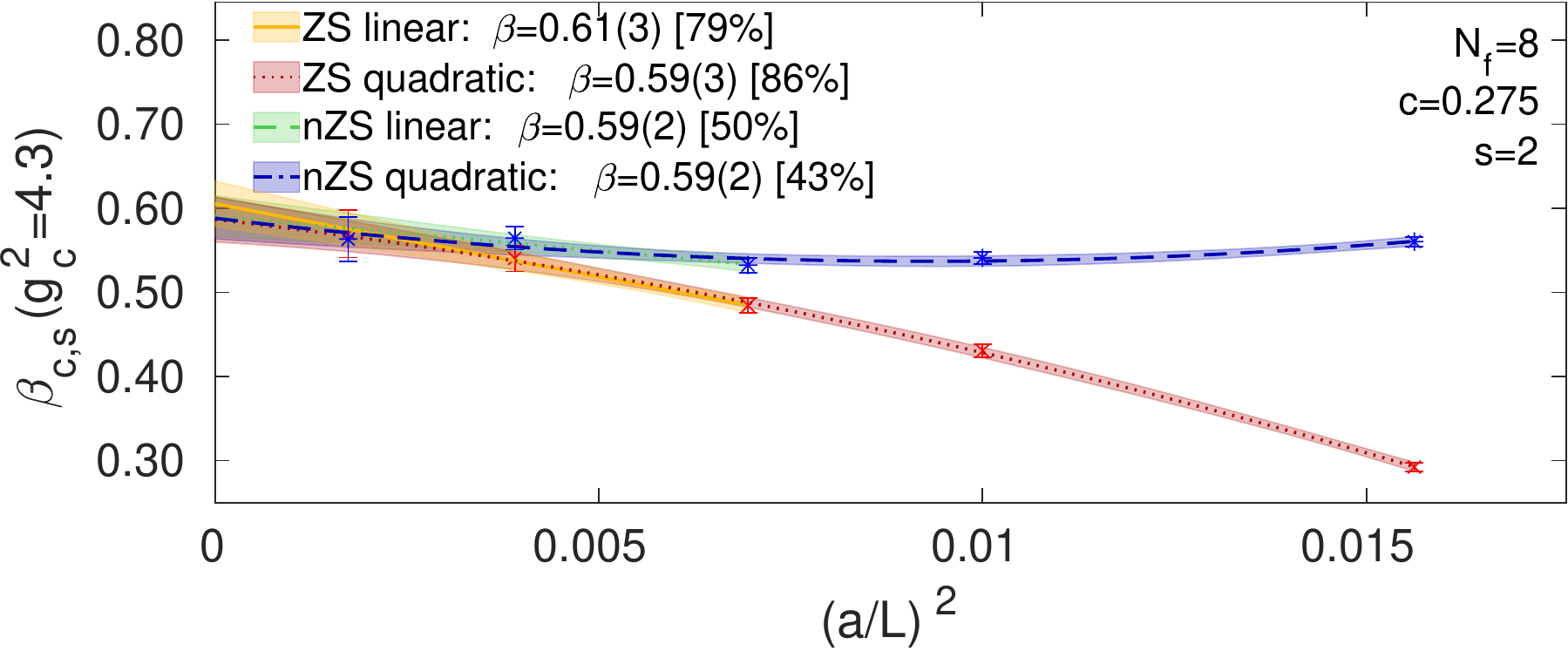}
  \end{minipage}
  \begin{minipage}{0.49\textwidth}
    \flushright
    \includegraphics[width=0.96\textwidth]{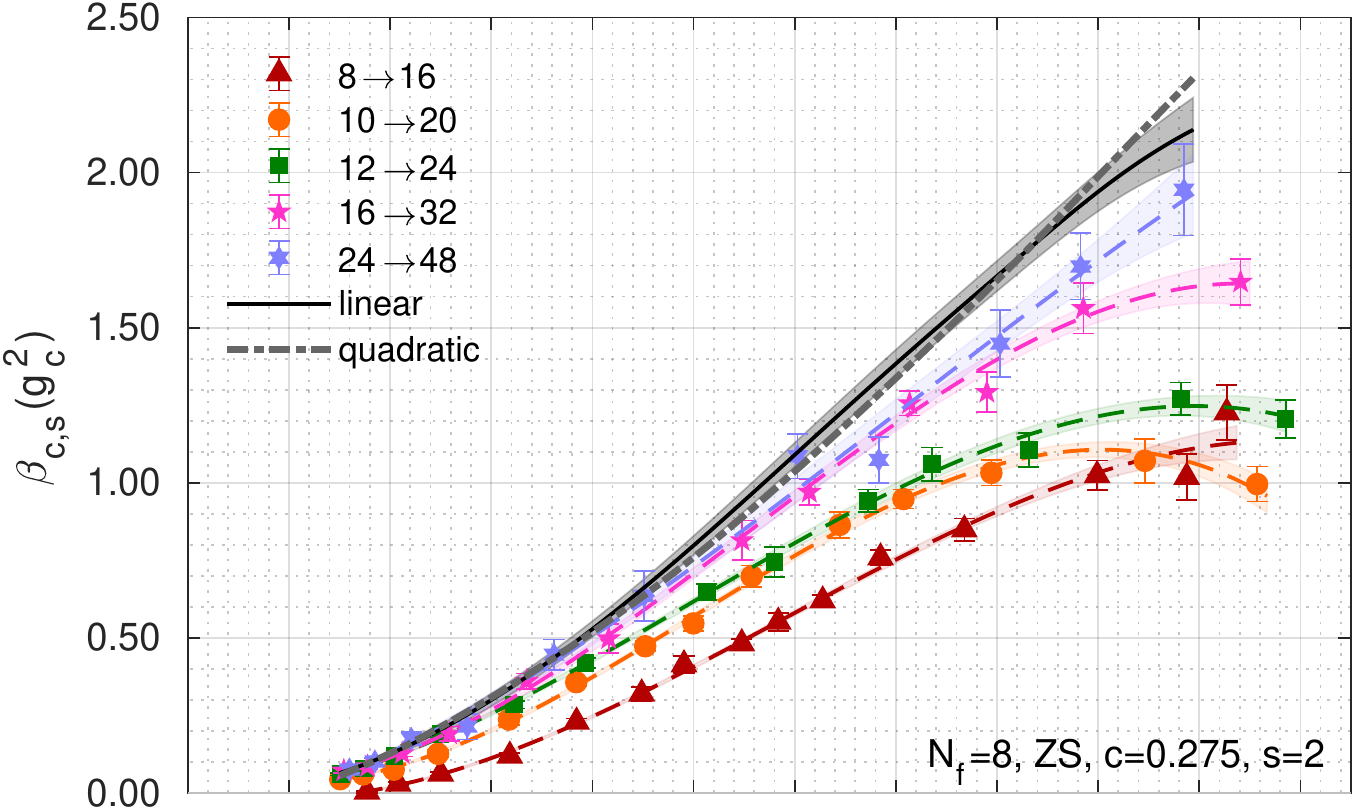}\\    
    \includegraphics[width=0.932\textwidth]{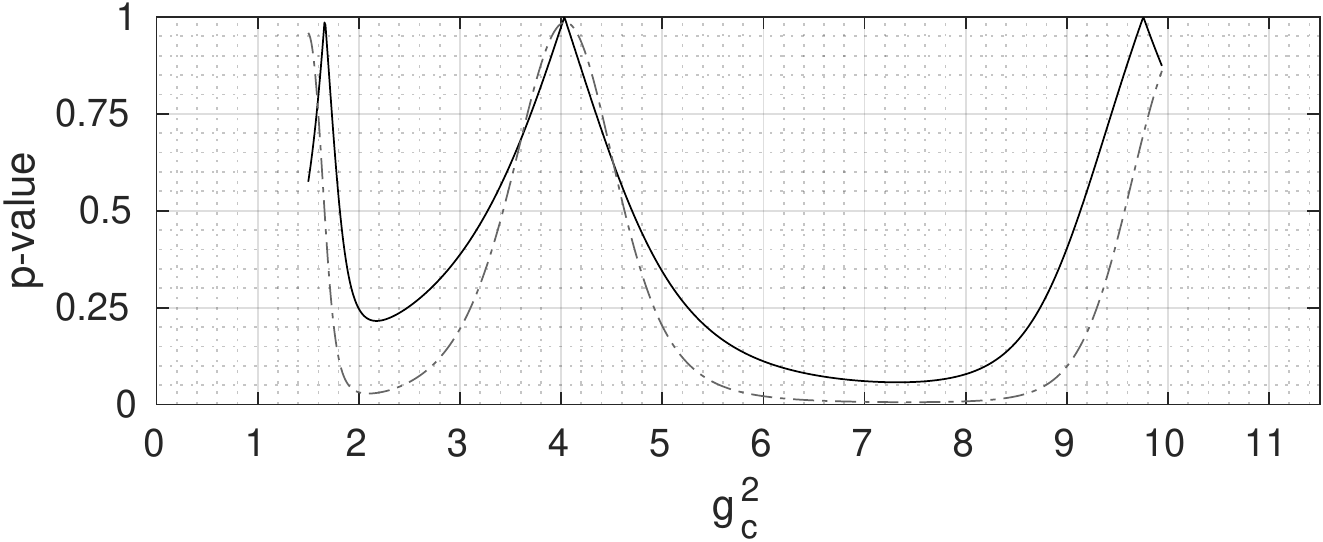} 
    \includegraphics[width=0.96\textwidth]{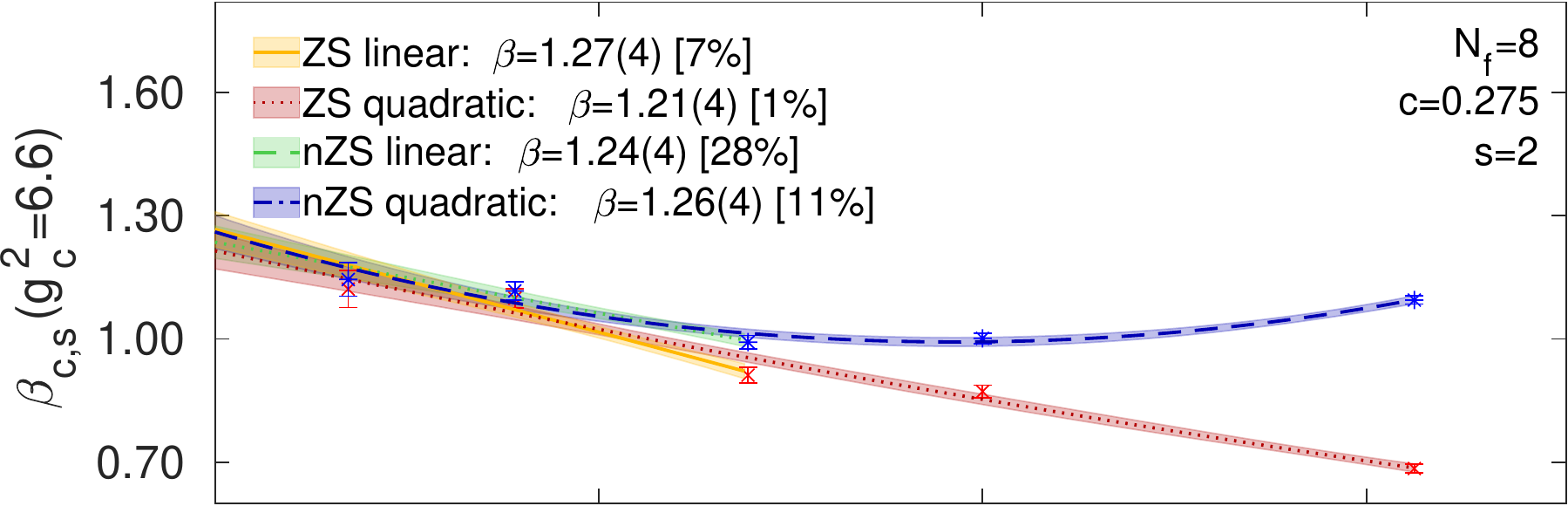}\\
    \includegraphics[width=0.96\textwidth]{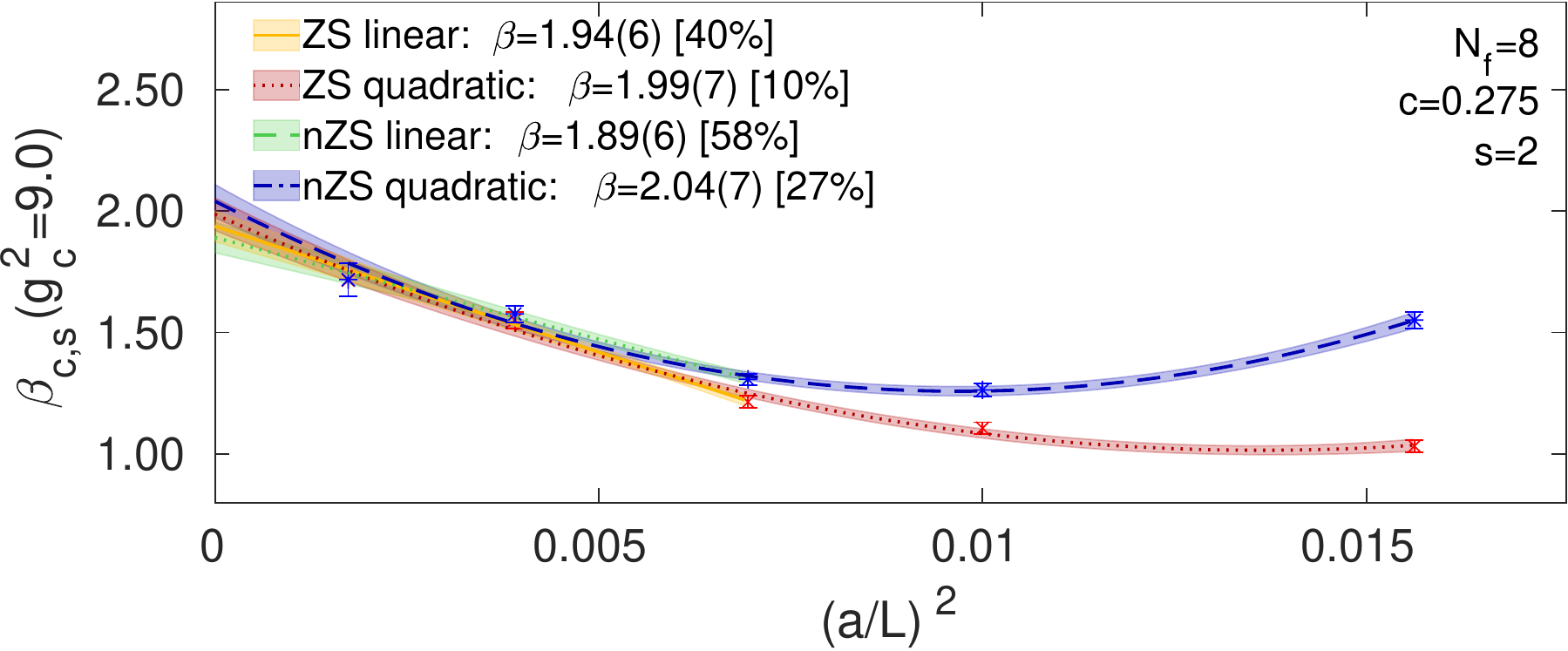}
  \end{minipage}
  \captionof{figure}{Discrete step-scaling $\beta$-function for $N_f=8$ in the $c=0.275$ gradient flow scheme for our preferred nZS (left) and ZS (right) data sets. The symbols in the top row show our results for the finite volume discrete $\beta$ function with scale change $s=2$. The dashed lines with shaded error bands in the same color of the data points show the interpolating fits. We consider two continuum limits: a linear fit (black line with gray error band) in $a^2/L^2$ to the three largest volume pairs and a quadratic fit to all volume pairs (black dash-dotted line). The $p$-values of the continuum extrapolation fits are shown in the plots in the second row. Further details of the continuum extrapolation at selected $g_c^2$ values are presented in the small panels at the bottom where the legend lists the extrapolated values in the continuum limit with $p$-values in brackets.}
  \label{Fig.Nf8_beta_c275}
\end{minipage}
\pagebreak

\begin{figure*}[t]
  \begin{minipage}{0.49\textwidth}
   \flushright 
   \includegraphics[width=0.96\textwidth]{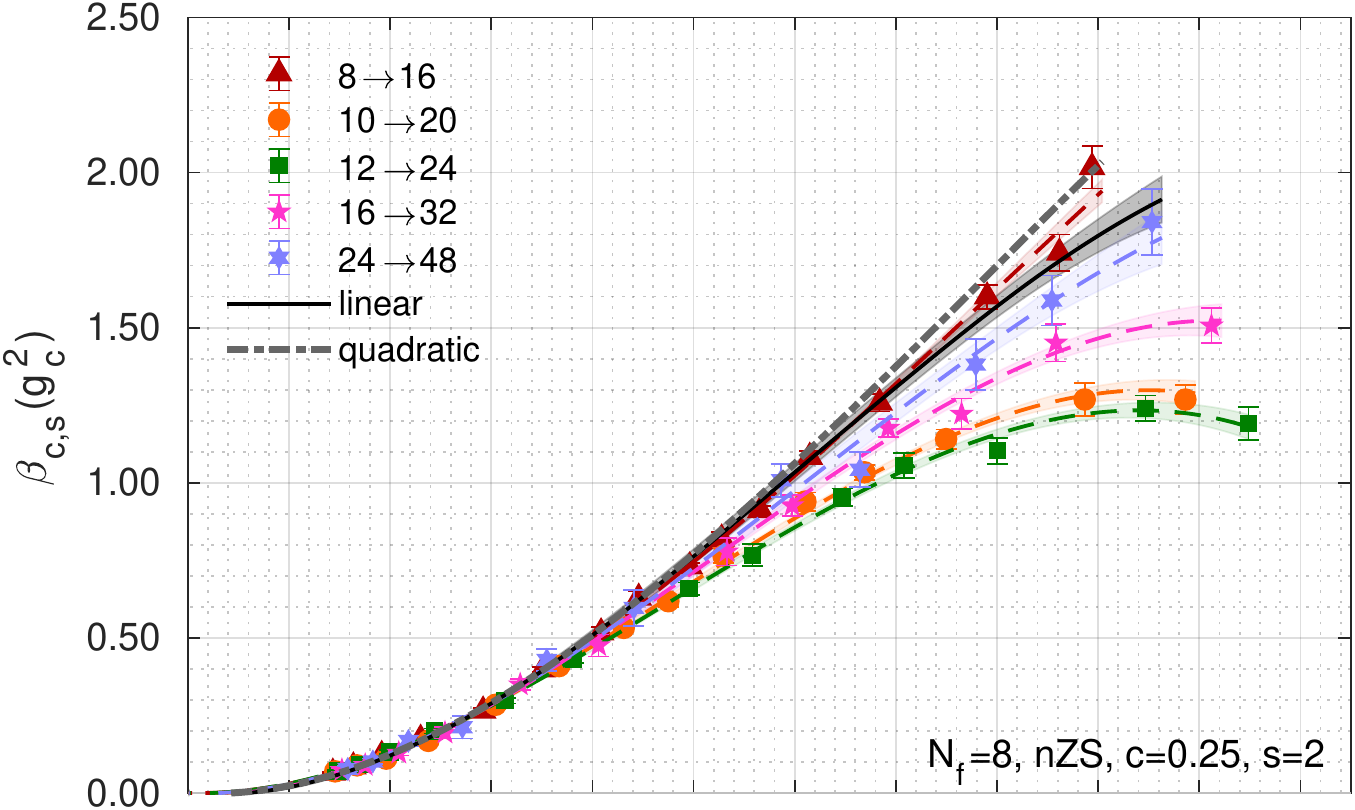}\\
   \includegraphics[width=0.932\textwidth]{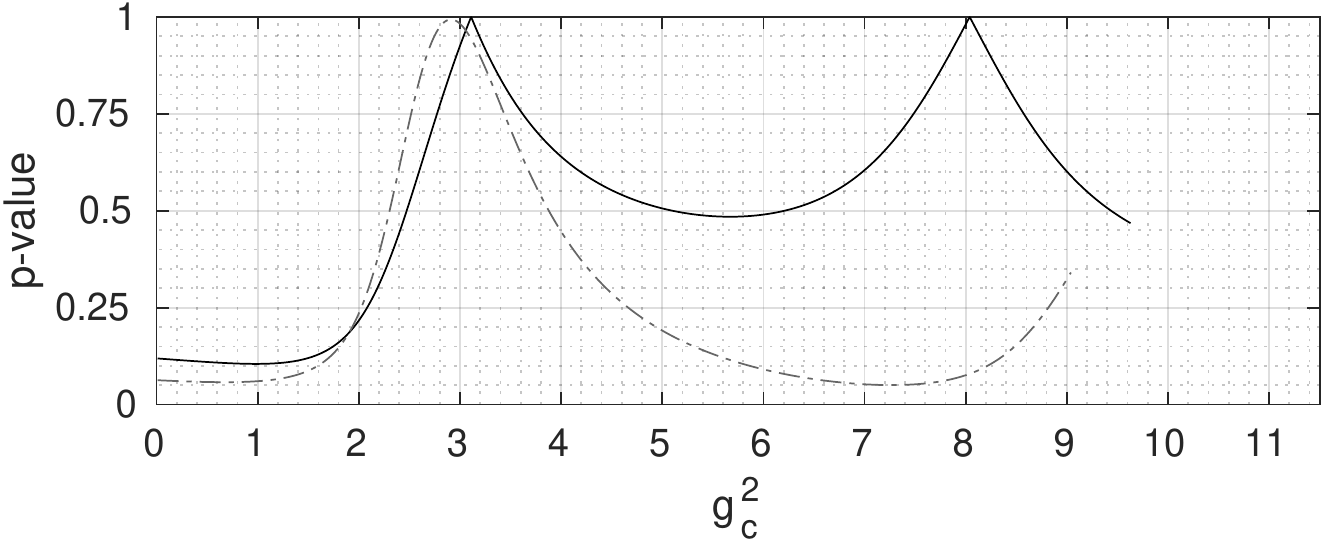} 
   \includegraphics[width=0.96\textwidth]{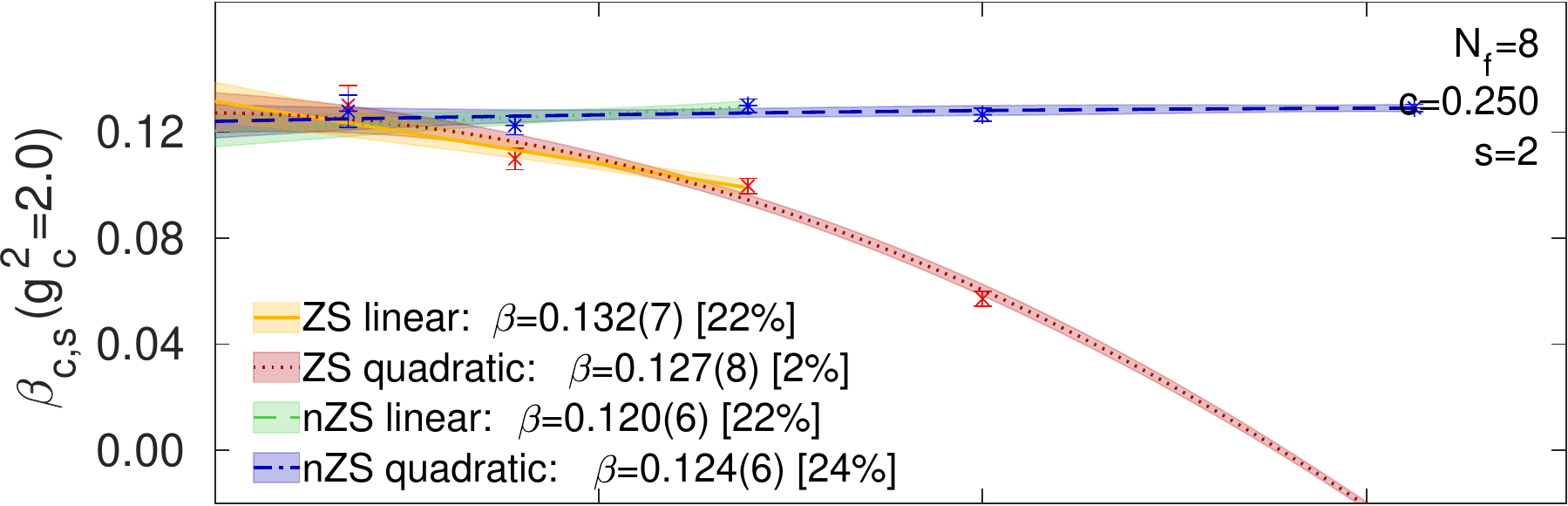}\\
   \includegraphics[width=0.96\textwidth]{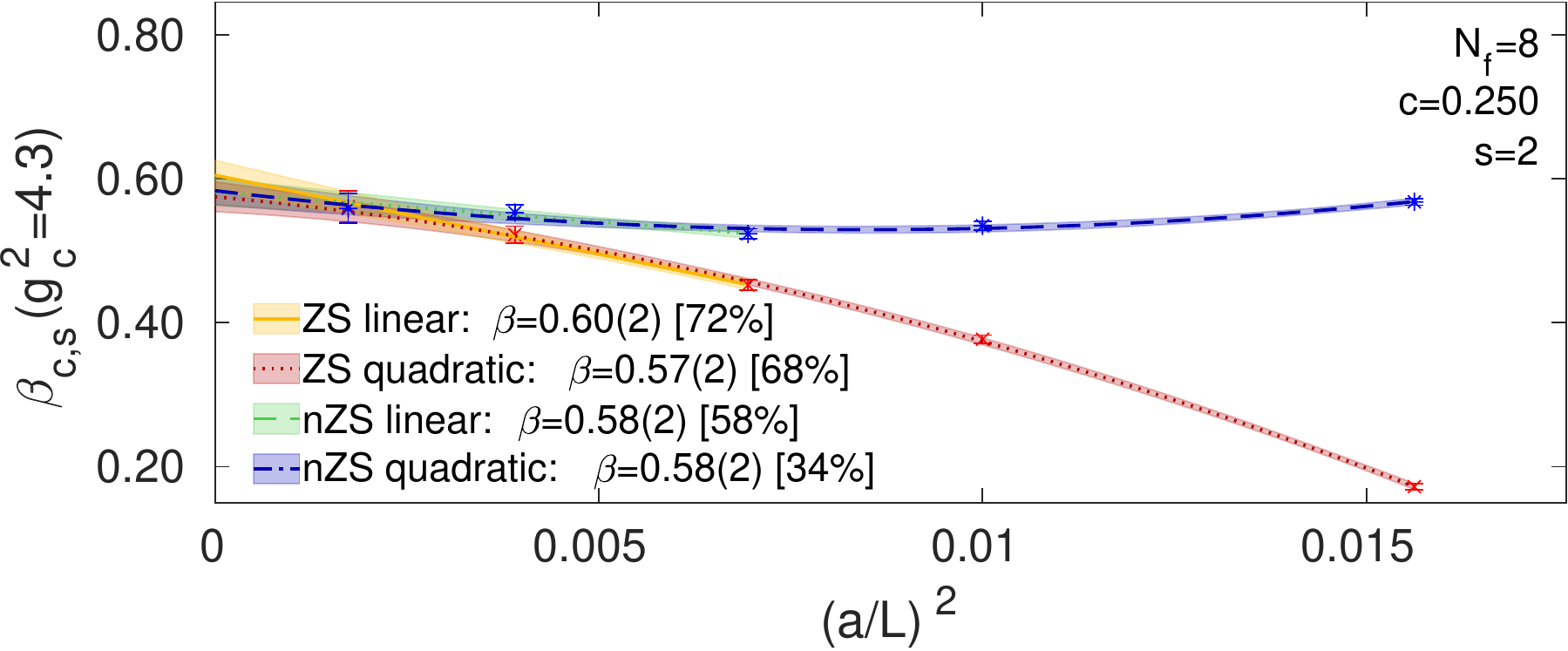}
  \end{minipage}
  \begin{minipage}{0.49\textwidth}
    \flushright
    \includegraphics[width=0.96\textwidth]{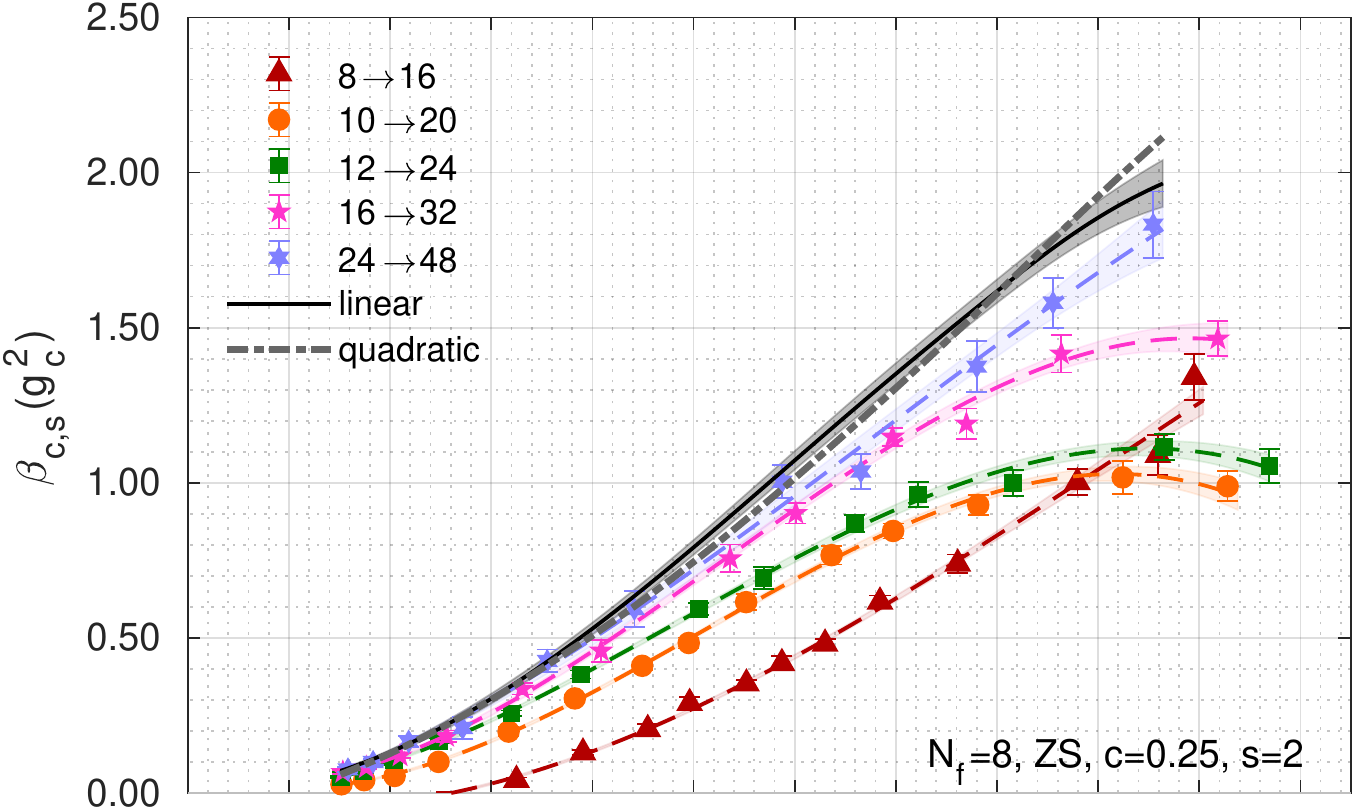}\\    
    \includegraphics[width=0.932\textwidth]{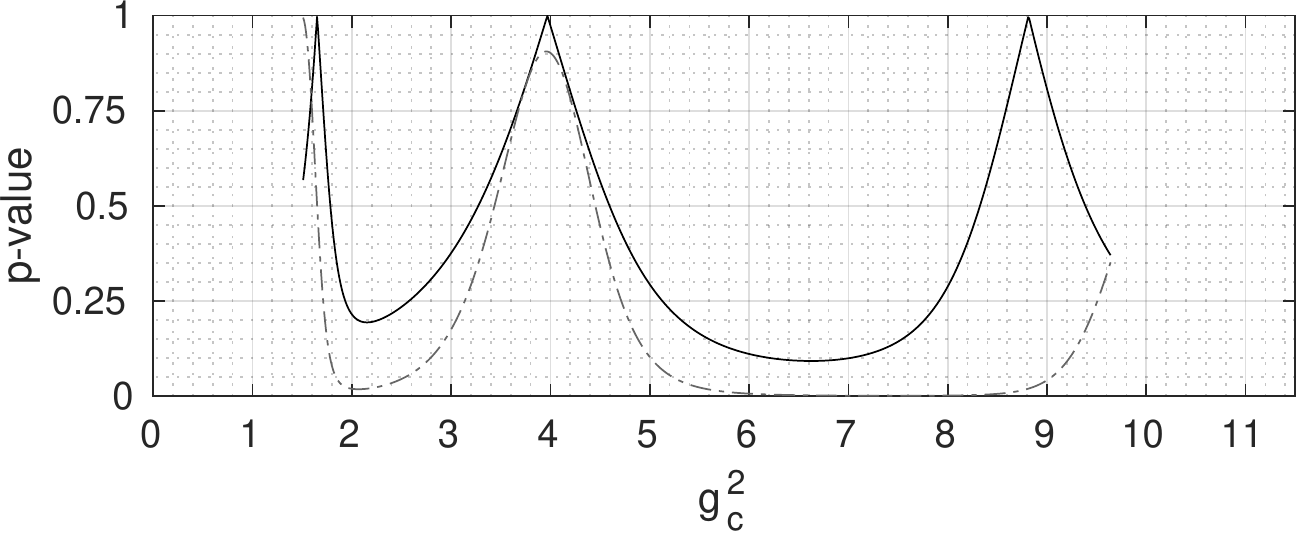} 
    \includegraphics[width=0.96\textwidth]{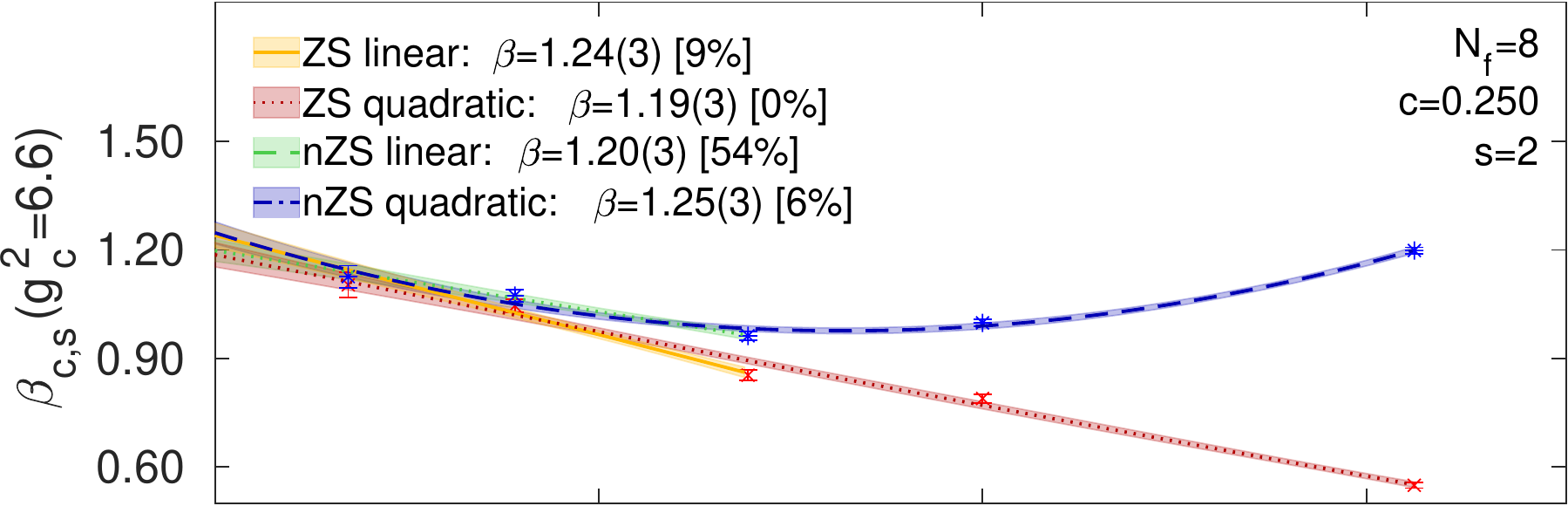}\\
    \includegraphics[width=0.96\textwidth]{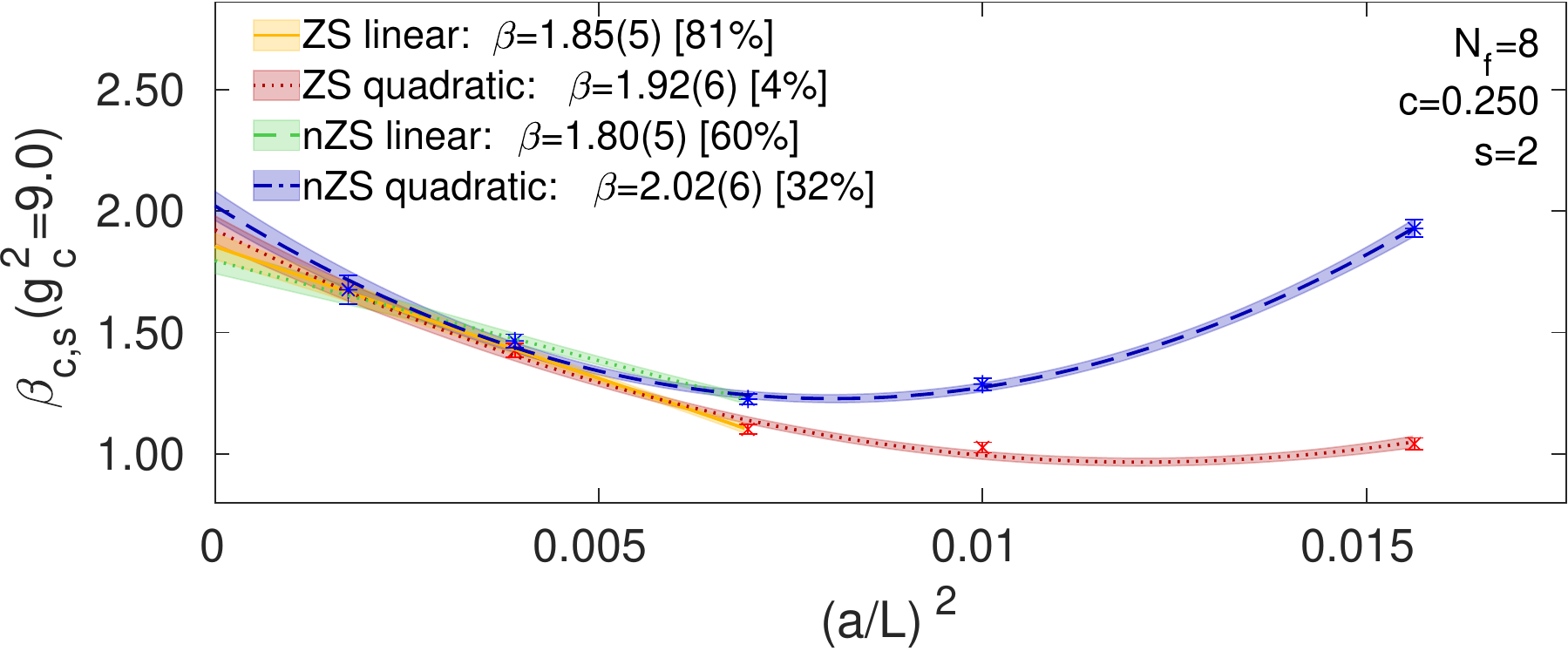}
  \end{minipage}
  \caption{Discrete step-scaling $\beta$-function for $N_f=8$ in the $c=0.250$ gradient flow scheme for our preferred nZS (left) and ZS (right) data sets. The symbols in the top row show our results for the finite volume discrete $\beta$ function with scale change $s=2$. The dashed lines with shaded error bands in the same color of the data points show the interpolating fits. We consider two continuum limits: a linear fit (black line with gray error band) in $a^2/L^2$ to the three largest volume pairs and a quadratic fit to all volume pairs (black dash-dotted line). The $p$-values of the continuum extrapolation fits are shown in the plots in the second row. Further details of the continuum extrapolation at selected $g_c^2$ values are presented in the small panels at the bottom where the legend lists the extrapolated values in the continuum limit with $p$-values in brackets. Only statistical errors are shown.}   
  \label{Fig.Nf8_beta_c250}
\end{figure*}

\bibliography{../General/BSM}
\end{document}